\documentclass[12pt]{article}
\usepackage{graphics}
\usepackage{graphicx}
\usepackage{amssymb}
\usepackage{amsmath}
\usepackage{hyperref}
\usepackage{enumitem}
\numberwithin{equation}{section}
\begin{document}
\title{\bf Relevant perturbation of entanglement entropy of singular surfaces }
\author{Mostafa Ghasemi\thanks{Email: ghasemi.mg@gmail.com}  \hspace{2mm} and
        Shahrokh Parvizi\thanks{Corresponding author: Email: parvizi@modares.ac.ir}\hspace{1mm}\\
		{\small {\em  Department of Physics, School of Sciences,}}\\
        {\small {\em Tarbiat Modares University, P.O.Box 14155-4838, Tehran, Iran}}\\
       }
\date{\today}
\maketitle

\abstract We study the entanglement entropy of theories that are derived from relevant perturbation of given CFTs for regions with a singular boundary by using the AdS/CFT correspondence. In the smooth case, it is well known that a relevant deformation of the boundary theory by the relevant operator with scaling dimension $\Delta=\frac{d+2}{2}$ generates a logarithmic universal term to the entanglement entropy. As the smooth case, when the boundary CFT deformed by a relevant operator, we find that the entanglement entropy of singular surface also contains a new logarithmic term which is due to relevant perturbation of the conformal field theory, and depends on the scaling dimension of relevant operator. We also find for extended singular surfaces, $c_{n}\times R^{m}$, as well as logarithmic term, the new universal double logarithmic terms may appear depending on the scaling dimension of relevant operator and spacetime dimensions. These new terms are due to relevant perturbation of the boundary theory.



\section{Introduction} \label{intro}
The entanglement structure of a quantum system can be quantified by the entanglement entropy which is applied in diverse ranging quantum systems  \cite{Ref1,Ref2,Ref3,Ref4,Ref5,Ref6,Ref7,Ref8,Ref9,Ref10,Ref11,Ref12,Ref13,Ref14}. In quantum field theory, it is defined for a spatial region ${A}$ as Von Neumann entropy of reduced density matrix $\rho_{A}$, $ S=-Tr( \rho_{A} log\rho_{A})$, in which the reduced density matrix defined with tracing out the degrees of freedom of the complementary region $\bar{A}$ of $ {A}$, $\rho_{A}=Tr_{\bar{A}}  (\rho)$.
The boundary of $A$ is called the entanglement surface $\Sigma=\partial{A}$.
 In the vacuum state of a $d$- dimensional CFT, the entanglement entropy for a smooth entangling surface takes the form \cite{Ref15}

\begin{equation}
    S_{EE}=c_{d-2} \frac{R^{d-2}}{\delta^{d-2}}+c_{d-4} \frac{R^{d-2}}{\delta^{d-4}}+...+
    \begin{cases}
    c_{1}\frac{R}{\delta}+(-1)^{\frac{d-1}{2}}s(\varSigma)_{univ}+... &  \text{d odd} \\
   c_{2}\frac{R^{2}}{\delta^{2}}+(-1)^{\frac{d-1}{2}}s(\varSigma)_{univ}\log\frac{R}{\delta} +... &  \text{d even}
    \end{cases}
\end{equation}
where $R$ is the characteristic size of the entangling surface $\varSigma$, $c_{i}$'s are the scheme dependent coefficients, and the leading divergence is the well-known area law term \cite{Ref16,Ref17}. $s(\varSigma)_{univ}$ is $R$- independent constant and gives the universal part of the entanglement entropy. But, in general, it depends on the shape of the entangling surface. In even dimensions, $s(\varSigma)_{univ}$ appears as coefficient of logarithmic term and is universal in the sense that is independent of regularization scheme. Typical computation shows that the coefficient of logarithmic term is a linear combination of the central charges which appear in the trace anomaly of the CFT \cite{Ref10,Ref12,Ref18,Ref19,Ref20,Ref21,Ref22,Ref23,Ref24}.

\indent
In the case that there is a singularity in the entangling surface, the entanglement entropy contains an additional singular term which is universal. For example in $d=3$ dimension, the EE is given by
\begin{equation}
 S_{EE}=\beta\frac{H}{\delta}- a(\Omega)\log(\frac{H}{\delta})+O(1),
\end{equation}
in which $\Omega$ is the opening angle, $\beta$ is a non-universal constant, and $H$ is the size of the entangling surface. The appearance of a new logarithmic term is due to corner shape of the entangling surface and the coefficient $a(\Omega)$ gives the universal part of the EE \cite{Ref25,Ref26,Ref27,Ref28,Ref29,Ref30,Ref31}. Similar terms appear in higher dimensional entangling singular surfaces \cite{Ref32,Ref33}\footnote{There are other related works in other background and context\cite{Ref44,Ref45,Ref46,Ref47,Ref48,Ref49,Ref50}.}.

The importance of these universal terms is that their coefficients encode the universal data of the underlying quantum field theory. 

One of the nice approaches to computation of entanglement entropy is laying in holographic context. According to holographic prescription \cite{Ref11,Ref12}, the entanglement entropy of a sub-region $A$ in boundary theory is given by
\begin{equation}
S_{EE}= \frac{Area(\gamma)}{4G_{N}}
\end{equation}
where $\gamma$ is the bulk minimal surface which is homologous to that of V and $\partial{\gamma}$ matches the entangling surface $\partial{A}$ in the boundary. The above formula holds in the case that the bulk physics is described by the Einstein gravity.

There are similar stories on the EE of the relevant perturbed conformal field theories \cite{Ref15,Ref34,Ref35,Ref36,Ref37,Ref38,Ref39,Ref40,Ref41}. As is well known, relevant perturbation of a conformal field theory induces a universal logarithmic term in the entanglement entropy, either from the field theoretic calculations in \cite{Ref34,Ref36,Ref37,Ref38} or the holographic computations.

In holography, the dual picture of a relevant perturbation corresponds to a massive scalar field in the bulk which can deform the background from a pure $AdS$ space to an asymptotically $AdS$. The scaling dimension of the relevant operator is related to mass of the scalar field. It has been shown in \cite{Ref35} that in the first order of perturbation only for dimension $\Delta=(d+2)/2$ the $S_{EE}$ receives a universal logarithmic term proportional to the scaling parameter. It is interesting to study the effect of relevant perturbation on the entanglement entropy of a singular region. In \cite{Ref41}, we considered a 3 dimensional CFT on the boundary of an asymptotically $AdS$ space which is perturbed by a massive scalar field and derived the entanglement entropy for a singular entangling surface by the Ryu-Takayanagi prescription \cite{Ref10,Ref11}. We found two independent universal logarithmic terms: One for the corner contribution to $S_{EE}$ and the other due to the relevant perturbation at dimension $\Delta=(d+2)/2$.

In this article, we extend our previous work \cite{Ref41} and study the effect of relevant perturbation of CFT on the entanglement entropy of higher dimensional singular regions including $c_{n}$, $k\times R^{m}$ and $c_{n}\times R^{m}$, which is possible when either the intrinsic or extrinsic curvatures have singularities \cite{Ref32}. In considering these geometries, we suppose that the background geometry is flat $R^{d}$, and write the metric in Euclidean signature as
\begin{equation}
ds ^{2}= -dt^{2}+d\rho^{2}+\rho^{2}(d\theta^{2}+\sin^{2}\theta  d\Omega_{n}^{2})+\sum_{i=1}^{m}(dx^{i})^{2},
\end{equation}
where $d\Omega_{n}^{2}$ denotes the metric on a unit sphere $S^n$, and $d=3+n+m$.
The entanglement entropy of a conformal field theory in vacuum contains a universal term for the singular entangling surface. Here we consider the holographic entanglement entropy of a relevant perturbation of a conformal field theory for these geometries, and identify that for $d$- dimensional relevant perturbed conformal field theories there are new universal logarithmic terms which depend on choosing the scaling dimension of relevant operator.

By the AdS/CFT correspondence, the relevant perturbation of boundary theory, with scaling dimension $\Delta<d$, is described by turning on a scalar degree of freedom in the bulk. Then the geometry is asymptotically AdS in the presence of scalar field \cite{Ref35,Ref42,Ref43}. So, we use the following bulk metric,
      \begin{equation}\label{si}
       ds ^{2}= \frac{L^{2}}{z^{2}}(-dt^{2}+d\rho^{2}+\rho^{2}(d\theta^{2}+\sin^{2}\theta  d\Omega_{n}^{2})+dz^{2}+\sum_{i=1}^{m}(dx^{i})^{2}),
      \end{equation}
for pure AdS spacetime, while for asymptotic AdS spacetimes  the following bulk metrics ansatz
\begin{equation}
 ds ^{2}= \frac{L^{2}}{z^{2}}(-dt^{2}+d\rho^{2}+\rho^{2}(d\theta^{2}+\sin^{2}\theta  d\Omega_{n}^{2})+\frac{dz^{2}}{f(z)}+\sum_{i=1}^{m}(dx^{i})^{2}),
\end{equation}
where $f(z)\rightarrow 1$ as $z\rightarrow 0$. For a source deformation and near the boundary, $f(z)$ can be expanded as
\begin{equation}\label{fz}
  f(z)= 1+\mu^{2\alpha}z^{2\alpha}+\cdots, \qquad z \rightarrow 0
\end{equation}
where $\mu$ is some mass scale\footnote{Since the only dimensionful parameter is the coupling $\lambda$ of the relevant operator, the dimensional analysis yields  $\mu\sim\lambda^{1/(d-\Delta)}$ \cite{Ref35,Ref39}. }, and $\alpha$ is a positive constant where for a source deformation, we have $\alpha= d-\Delta$ in the standard quantization $d/2<\Delta<d$, and $ \alpha=\Delta$ in the alternative quantization $d/2-1< \Delta <d/2$. $\Delta$ is the scaling dimension of the relevant operator\footnote{In the case for which $\Delta=d/2$, we should replace $\mu^{2\alpha}z^{2\alpha}$ in \eqref{fz} by $ (\mu z)^{d}(\log(\mu z))^{2}$.}.

This paper is organized as follows. In section 2, we review the holographic entanglement entropy of geometries in the form  $c_{n}$, $k\times R^{m}$ and $c_{n}\times R^{m}$  in AdS background. In section 3, we derive holographic entanglement entropy of these geometries in AAdS backgrounds which corresponds to the relevant perturbation of the CFT. In the last section, we discuss and conclude our results.


\section{Holographic Entanglement Entropy of Crease in pure AdS background}
\label{sec3}
In this section, we will review the entanglement entropy of the singular regions in the form  $c_{n}$, $k\times R^{m}$ and $c_{n}\times R^{m}$ for holographic CFTs that dual to Einstein gravity \cite{Ref32}. First we review the entanglement entropy of creases $c_{n}\times R^{m}$, then we consider $k\times R^{m}$ and cone $c_{n}$.

\subsection{Crease $c_{n}\times R^{m}$  }
In this subsection, we review the holographic entanglement entropy for singular surfaces in the form of $c_{n}\times R^{m}$ in the anti-de Sitter space time background in arbitrary dimension. The metric is given in \eqref{si}. The geometry of crease $c_{n}\times R^{m}$ is defined by $\theta\in[0,\frac{\Omega}{2}]$ and $\rho\in[0,\infty]$. We denote the coordinates over the bulk minimal surface by $\sigma=(z,\theta,\xi^{i},x^{j})$, where $\xi^{i}$'s are coordinates on sphere $S^{n}$, and $x^{j}$'s are on $R^{m}$. Due to rotational symmetry along the sphere $S^{n}$, we parameterize the bulk minimal surface as $\rho=\rho(z,\theta)$. So the induced metric on the entangling surface in time slice $t=0$ becomes
\begin{align}
 ds ^{2}&=\gamma_{ab}dx^adx^b  \nonumber\\
&=\frac{L^{2}}{z^{2}}\Big( \big(\rho^{'2}+1\big)dz^{2}+2\rho'\dot{\rho}dzd\theta+(\dot{\rho}^{2}+\rho^{2})
 d\theta^{2}+\rho^{2}\sin^{2}\theta  d\Omega_{n}^{2}+\sum_{i=1}^{m}(dx^{i})^{2}),
\end{align}

\begin{equation*}
\mathbf{\gamma_{ij}} = \left(
\begin{array}{cccccc}
\frac{L^{2}}{z^{2}}\big(\rho^{'2}+1\big) & \frac{L^{2}}{z^{2}}\rho'\dot{\rho}&& \\
\frac{L^{2}}{z^{2}}\rho'\dot{\rho} &\frac{L^{2}}{z^{2}}(\dot{\rho}^{2}+\rho^{2}) & & \\
 & & &\\
  &  &\frac{L^{2}}{z^{2}}\sin^{2}g_{ab}(S^{n})&   \\
  & & &\frac{L^{2}}{z^{2}}&   \\
 & & & &\ddots\\
 & && &&\frac{L^{2}}{z^{2}}
\end{array} \right)
\end{equation*}
so that
\begin{equation}\label{gamma}
  \sqrt{\gamma}=\frac{L^{d-1}}{z^{d-1}}\rho^{n}\sin^{n}\theta\sqrt{\big(\rho^{'2}+1\big)\rho^{2}+\dot{\rho}^{2}},
\end{equation}
By the RT prescription, the entanglement entropy is derived as
\begin{equation}
S_{EE}=\frac{1}{4G_{N}}\int d\sigma \sqrt{\gamma} = \frac{L^{d-1}\Omega_{n}\tilde{H}^{m}}{4G_{N}}\int dzd\theta   \frac{\rho^{n}\sin^{n}\theta}{z^{d-1}}\sqrt{(\rho^{'2}+1)\rho^{2}+\dot{\rho}^{2}}
\end{equation}
where $\Omega_{d-3}$ is the area of $(d-3)-$sphere, $\tilde{H}^{m}$ is volume of $m$-dimensional space, $\dot{\rho}=\partial_{\theta}\rho$ and $\rho'=\partial_{z}\rho$. In the integration over  $x^{i}$'s we have used the $IR$ cut-off $x^{i}\in[-\frac{\tilde{H}}{2},\frac{\tilde{H}}{2}]$. From the entropy functional we can find that the equation of motion of $\rho(z,\theta)$ to be
\begin{align}
&z\sin\theta\rho^{2}(\rho^{2}+\dot{\rho}^{2})\rho''
+z\sin\theta\rho^{2}(1+\rho^{'2})\ddot{\rho}
-2z\sin\theta\rho^{2} \rho' \dot{\rho}\dot{\rho}'+nz\cos\theta\dot{\rho}((1+\rho^{'2})\rho^{2}+\dot{\rho}^{2})
\nonumber\\
&-z\sin\theta\rho((n+1)(1+\rho^{'2})\rho^{2}+(n+2)\dot{\rho}^{2})
-(d-1)\sin\theta\rho^{2}\rho'((1+\rho^{'2})\rho^{2}+\dot{\rho}^{2})=0
\end{align}
where $\ddot{\rho}=\partial_{\theta}^{2}\rho$, $\rho''=\partial_{z}^{2}\rho$ and $\dot{\rho}'=\partial_{\theta}\partial_{z}\rho$.
 Once again, due to scaling symmetry we make the following ansatz:
\begin{equation}
  \rho(z,\theta)=\frac{z}{h(\theta)}
\end{equation}
so the entropy functional takes the form
\begin{align}\label{eq 4}
&S_{EE}=\frac{L^{d-1}\Omega_{n}\tilde{H}^{m}}{2G_{N}}\int_{\delta}^{z_{m}}dz \int_{0}^{\frac{\Omega}{2}-\epsilon}d\theta
\Big[\frac{\sin^{n}(\theta)\rho^{n}\sqrt{\rho^{2}(1+\rho^{'2})+\dot{\rho}^{2}}}{z^{d-1}}\Big]
\nonumber\\
 &\qquad =\frac{L^{d-1}\Omega_{n}\tilde{H}^{m}}{2G_{N}}\int_{\delta}^{z_{m}}\frac{dz}{z^{d-n-2}} \int_{h_{0}}^{h_{c}}dh
\frac{\sin^{n}(\theta)\sqrt{1+h^{2}+\dot{h}^{2}}}{\dot{h}h^{n+2}}
\end{align}
 in which, $h_{0}=h(0)$, $z=\delta$ is $UV$ cut-off, $h_{c}=h(\frac{\Omega}{2}-\epsilon)$, $\dot{h}=\partial_{\theta}h$, and $\dot{h_{0}}=0$. Note, in the second line we have changed the integration over $\theta$ to the integration over $h$. Using this entropy functional, we find the equation of motion for $h$
\begin{align}
&\sin(\theta) h(1+h^{2})\ddot{h}+\sin(\theta)(nh^{2}+d-1)\dot{h}^{2}
\nonumber\\
&+n\cos(\theta)h\dot{h}(1+h^{2}+\dot{h}^{2})+\sin(\theta)(1+h^{2})(d-1+(n+1)h^{2})=0,
\end{align}

In order to identify various divergence structure that may be appeared, we must find the asymptotic behavior of integrand of \eqref{eq 4} in terms of $h$, where  $h\rightarrow0$. To do so, we make a change of variable $y=\sin(\theta)$ and independent variable from $\theta$ to $h$, and find the equation of motion of $y=y(h)$. Using the relations $\dot{h}=\frac{\sqrt{1-y^{2}}}{\dot{y}(h)}$, $\ddot{h}=-\frac{y\dot{y}^{2}+(1-y^{2})\ddot{y}}{\dot{y}^{3}}$, we reach to the following equation
\begin{align}
&h(1+h^{2})y(1-y^{2})\ddot{y}-(1+h^{2})(d-1+(n+1)h^{2})y\dot{y}^{3}+(1+h^{2})h((1+n)y^{2}-n)\dot{y}^{2}
\nonumber\\
&-nh(1-y^{2})^{2}-(nh^{2}+d-1)y(1-y^{2})\dot{y}=0
\end{align}
where $\dot{y}(h)=\frac{dy}{dh}$ and $\ddot{y}(h)=\frac{d^{2}y}{dh^{2}}$. Now, by solving  this equation perturbatively in terms of $h$ near the boundary, where $h$ is small with boundary condition $y=\sin(\frac{\Omega}{2})$ at $h=0$, we find that
\begin{align}
&y=\sin(\frac{\Omega}{2})+\frac{n\cos(\frac{\Omega}{2})\cot(\frac{\Omega}{2})}{4-2d}h^{2}
\nonumber\\
&-\frac{n\csc^{5}(\frac{\Omega}{2})\Big[\Big((d-2)^{2}-2n\Big)n+\Big(2(d-2)^{2}-(d-2)dn+2n^{2}\Big)\sin^{2}(\frac{\Omega}{2})\Big]\sin^{2}(\Omega)}{32(d-4)(d-2)^{3}}h^{4}+...
\end{align}
Of course, this should be modified in $d=4$ and other even dimensions for which we generate the solution for $y$ in the Appendix \ref{A} and show that it includes logarithmic terms.
Near the boundary, the integrands of entropy functional \eqref{eq 4} in the asymptotic limit behaves as\footnote{as explained in the appendices, in even dimensions, the logarithmic terms in the $y$ solution induce some logarithmic terms in the following expansion. However, they are less singular and do not contribute to the entanglement entropy. These logarithmic terms would be important in the relevant perturbation of the theory in the next section.}
\begin{align}\label{expansion1}
&\frac{\sin^{n}(\theta)\sqrt{1+h^{2}+\dot{h}^{2}}}{\dot{h}h^{n+2}}\sim-\frac{\sin^{n}(\frac{\Omega}{2})}{h^{n+2}}+P_{n}\frac{1}{h^{n}}+P_{n-2}\frac{1}{h^{n-2}}+P_{n-4}\frac{1}{h^{n-4}}+\cdots,
\end{align}
where a few coefficients $P$'s are found in the Appendix \ref{B}. Substitute this expansion in the entropy functional \eqref{eq 4}:
\begin{equation}
S_{EE}=\frac{L^{d-1}}{2G_{N}}\Omega_{n}\tilde{H}^{m}\Big(I_{1}+I_{2} \Big),
\end{equation}
where $I_{1}$ and $I_{2}$ defined as below:
\begin{align}
&I_{1}=\int_{\delta}^{z_{m}}\frac{dz}{z^{d-n-2}} \int_{h_{0}}^{h_{c}}dh
\Big[\sin^{n}(\theta)\frac{\sqrt{1+h^{2}+\dot{h}^{2}}}{\dot{h}h^{n+2}}+\frac{\sin^{n}(\frac{\Omega}{2})}{h^{n+2}}-\sum_{i=0}\frac{P_{(n-2i)}}{h^{n-2i}}\Big],
\end{align}
and
\begin{align}
&I_{2}=\int_{\delta}^{z_{m}}\frac{dz}{z^{d-n-2}} \int_{h_{0}}^{h_{c}}dh
\Big[-\frac{\sin^{n}(\frac{\Omega}{2})}{h^{n+2}}+\sum_{i=0}P_{(n-2i)}\frac{1}{h^{n-2i}}\Big].
\end{align}
We differentiate these terms with respect to $UV$ cut-off  $\delta$ and look for various divergent terms. With some manipulation, we find that
\begin{align}
\frac{dI_{1}}{d\delta}&=
-\frac{1}{\delta^{d-n-2}} \int_{h_{0}}^{0}dh
\Big[\frac{\sin^{n}(\theta)\sqrt{1+h^{2}+\dot{h}^{2}}}{\dot{h}h^{n+2}}+\frac{\sin^{n}(\frac{\Omega}{2})}{h^{n+2}}-\sum_{i=0}P_{(n-2i)}\frac{1}{h^{n-2i}}\Big]
\nonumber\\
&=-\frac{1}{\delta^{d-n-2}} \int_{h_{0}}^{0}dh J_{1}(h).
\end{align}
where we define:
\begin{align}
&J_{1}(h)=  \Big[\frac{\sin^{n}(\theta)\sqrt{1+h^{2}+\dot{h}^{2}}}{\dot{h}h^{n+2}}+\frac{\sin^{n}(\frac{\Omega}{2})}{h^{n+2}}-\sum_{i=0}P_{(n-2i)}\frac{1}{h^{n-2i}}\Big].
\end{align}
In order to single out $log$ terms in $I_{2}$, we note that in $dI_2/d\delta$, there are two contributions to powers of $\delta$ which are from $z$ and $h$ integrations, respectively. When both $d$ dimension and $n$ are even, there is always a $1/\delta$ in $dI_2/d\delta$ corresponding to $P_{(n-d+4)}=P_{(1-m)}$ term by choosing $2i=d-4$. Then we find the entanglement entropy to be,
\begin{align}\label{ScnRmeven}
S_{EE}=&\frac{L^{d-1}\Omega_{n}\tilde{H}^{m}}{2G_{N}}\left[\frac{\sin^{n}(\frac{\Omega}{2})}{(n+1)(d-2)}\frac{H^{n+1}}{\delta^{d-2
}}- \epsilon_d\frac{P_{(1-m)}}{mH^{m}}\log(\frac{\delta}{H }) \right.\nonumber\\
&\left.- \sum'_{i=0}\frac{P_{(n-2i)}}{(n-2i-1)(d-2i-4)}\frac{H^{n-2i-1}}{\delta^{d-2i-4
}}+ \Big(\int_{h_{0}}^{0}dh  J_{1}(h)+F(h_{0},\Omega)\Big)\frac{1}{m\delta^{m}}\right]
\end{align}
where we introduce the following parameters
\begin{align}\label{epsilon}
\epsilon_\beta&\equiv\frac{1}{2}(1+(-1)^\beta), \nonumber\\
\bar{\epsilon}_\beta&\equiv\frac{1}{2}(1-(-1)^\beta).
\end{align}
The prime over summation means excluding $2i=d-4$ and we have replaced $d-n=m+3$ and define:
\begin{align}
&F(h_{0},\Omega)= \frac{-\sin^{n}(\frac{\Omega}{2})}{n+1}\frac{1}{h_{0}^{n+2
}}+\sum_{i=0}\frac{P_{(n-2i)}}{n-2i-1}\frac{1}{h_{0}^{n-2i-1}}
\end{align}

As we see in (\ref{ScnRmeven}) a logarithmic term appears. This is not due to singularity of entangling surface but is due to the even dimension of spacetime \cite{Ref32}.

When $n$ is odd, irrespective  of $d$ be even or odd, a $1/\delta$ term appears in $I_2$ as follows
\begin{align}
&I_{2}=\int_{\delta}^{z_{m}}\frac{dz}{z^{d-n-2}} \int_{h_{0}}^{h_{c}}dh
\Big[-\frac{\sin^{n}(\frac{\Omega}{2})}{h^{n+2}}+\epsilon_{n}P_{1}\frac{1}{h}+\sum'_{i=0}P_{(n-2i)}\frac{1}{h^{n-2i}}\Big].
\end{align}
where the prime over summation indicates excluding $i=n-1$ term. Now differentiate these terms with respect to $UV$ cut-off  $\delta$, 
\begin{align}
\frac{dI_{2}}{d\delta}=&
-\frac{\sin^{n}(\frac{\Omega}{2})}{n+1}\frac{H^{n+1}}{\delta^{d-1
}}+\sum'_{i=0}\frac{P_{(n-2i)}}{n-2i-1}\frac{H^{n-2i-1}}{\delta^{d-2i-3
}}-\epsilon_{n}\frac{P_{1}}{\delta^{m+1}}\log(\frac{\delta}{H}),
\nonumber\\
&-\frac{1}{\delta^{m+1}} \Big(\int_{h_{0}}^{0}dh  J_{1}(h)+F(h_{0},\Omega)\Big)
\end{align}
Notice that as before the second term contains a $1/\delta$ term by choosing $i=d-4$ when $d$ is even.  In the third term, as long as $m>0$, the $1/\delta$ does not appear and we therefore do not expect a double log term in the entropy. Finally we find
\begin{align}
S_{EE}=&\frac{L^{d-1}\Omega_{n}\tilde{H}^{m}}{2G_{N}}\left[\frac{\sin^{n}(\frac{\Omega}{2})}{(n+1)(d-2)}\frac{H^{n+1}}{\delta^{d-2
}}+\frac{\epsilon_{n}P_{1}}{m\delta^{m}}\Big(\log(\frac{\delta}{H})-\frac{1}{m}\Big) \right.\nonumber\\
&\left.- \sum'_{i=0}\frac{P_{(n-2i)}}{(n-2i-1)(d-2i-4)}\frac{H^{n-2i-1}}{\delta^{d-2i-4
}}+\frac{1}{m\delta^{m}} \Big(\int_{h_{0}}^{0}dh  J_{1}(h)+F(h_{0},\Omega)\Big)
\right.\nonumber\\
&\left.- \epsilon_d\frac{ P_{(1-m)}}{mH^{m}}\log(\frac{\delta}{H })\right]
\end{align}
This generalizes our results for any $n$ and $d$. Again the logarithmic term appears in even dimension $d$. In conclusion, in any cases, odd or even dimensional flat locus, we do not have double log term due to singularity of entangling surface, instead, we find log term which is due to even dimension of spacetime \cite{Ref32}. As explicit example we can consider two cases $c_{2}\times R^{1}$ and $c_{1}\times R^{2}$ in $d=6$ dimensions. Explicit computations shows that
\begin{align}
&S_{c_{1}\times R^{2}}=\frac{L^{5}\pi\tilde{H}^{2}}{G_{N}}\Big[\frac{\sin(\frac{\Omega}{2})}{8}\frac{H^{2}}{\delta^{4
}}+\frac{3(13-19\cos(\Omega) )\cot^{2}(\frac{\Omega}{2})\csc(\frac{\Omega}{2})}{8192H^{2}}\log (\frac{\delta}{H})+\cdots\Big]
\end{align}
and
\begin{align}
&S_{c_{2}\times R^{1}}=\frac{L^{5}\pi\tilde{H}^{2}}{G_{N}}\Big[\frac{\sin(\frac{\Omega}{2})}{8}\frac{H^{2}}{\delta^{4
}}+\frac{2(7-9\cos(\Omega) )\cot^{2}(\frac{\Omega}{2})}{256H}\log (\frac{\delta}{H})+\cdots\Big]
\end{align}
As we see, the logarithmic term shows up.

In odd dimensional spacetime with odd or even flat dimensional locus, we do not have any log term, and the effect of singularity is reflected in divergence term in the form $\delta^{n+3-d}\log(\delta/H)$ \cite{Ref32}. As explicit example we can consider  $c_{1}\times R^{1}$ in $d=5$ dimensions,
\begin{align}
&S_{c_{1}\times R^{1}}=\frac{L^{4}\pi\tilde{H}}{G_{N}}\Big[\frac{\sin(\frac{\Omega}{2})}{6}\frac{H^{2}}{\delta^{3
}}+\frac{\cot(\frac{\Omega}{2})\cos(\frac{\Omega}{2})}{9}\frac{1}{\delta}\log (\frac{\delta}{H})+\cdots\Big]
\end{align}
Here, the logarithmic term due to adding flat locus disappears \cite{Ref32}. We must note that, as was shown in \cite{Ref32}, adding the curved locus, effect of singularity of entangling surface are exhibited in terms of log or double log terms, depending on even or odd dimensional spacetime as well as dimension of curved locus.


\subsection{Cone  $c_{n}$ }
Now let us review the holographic entanglement entropy of the cone geometry $c_{n}$ in $d$- dimensions. In following we focus on $d=4,5,6$. Recalling the metric \eqref{si} with setting $m=0$. The cone geometry is defined by $\theta\in[0,\frac{\Omega}{2}]$ and $\rho\in[0,H]$, where we introduce $H$ as the IR cut-off for the geometry. We parameterize the bulk minimal surface as $\rho=\rho(z,\theta)$. So the induced metric on entangling surface in time slice $t=0$ becomes
\begin{equation}
 ds ^{2}= \frac{L^{2}}{z^{2}}\Big( \big({\rho}^{'2}+1\big)dz^{2}+2\rho'\dot{\rho}dzd\theta+(\dot{\rho}^{2}+\rho^{2})
 d\theta^{2}+\rho^{2}\sin^{2}\theta  d\Omega_{n}^{2}\Big),
\end{equation}

In this case, the calculations goes the same as Eqs. (\ref{gamma}) to (\ref{expansion1}) in $c_n\times R^m$ case with $m=0$ and $n=d-3$. 

It can be shown that in any odd dimension $d$ we have a log term and in even $d$ we have a double logarithmic term in the entanglement entropy. In the following we show this explicitly. Let us start with
\begin{align}
I&=\int_{\delta}^{z_{m}}dz \int_{0}^{\frac{\Omega}{2}-\epsilon}d\theta
\Big[\frac{\sin^{n}(\theta)\rho_{0}^{n}\sqrt{\rho_{0}^{2}(1+\rho_{0}^{'2})+\dot{\rho_{0}}^{2}}}{z^{d-1}}\Big]
\nonumber\\
&=\int_{\delta}^{z_{m}}\frac{dz}{z^{d-n-2}} \int_{h_{0}}^{h_{c}}dh
\frac{\sin^{n}(\theta)\sqrt{1+h^{2}+\dot{h}^{2}}}{\dot{h}h^{n+2}}
\end{align}
where $n=d-3$ and $\rho(z,\theta)=z/h(\theta)$. In order to identify the divergence terms we expand the integrand near boundary in terms of $h$. We find that
\begin{align}
&\frac{\sin^{n}(\theta)\sqrt{1+h^{2}+\dot{h}^{2}}}{\dot{h}h^{n+2}}\sim-\frac{\sin^{n}(\frac{\Omega}{2})}{h^{n+2}}+P_{n}\frac{1}{h^{n}}+P_{n-2}\frac{1}{h^{n-2}}+P_{n-4}\frac{1}{h^{n-4}}+\cdots,
\end{align}
Substitute this expansion in the entropy functional:
\begin{align}
I&=\int_{\delta}^{z_{m}}\frac{dz}{z^{d-n-2}} \int_{h_{0}}^{h_{c}}dh
\frac{\sin^{n}(\theta)\sqrt{1+h^{2}+\dot{h}^{2}}}{\dot{h}h^{n+2}}
\nonumber\\
&=\int_{\delta}^{z_{m}}\frac{dz}{z^{d-n-2}} \int_{h_{0}}^{h_{c}}dh
\Big[\sin^{n}(\theta)\frac{\sqrt{1+h^{2}+\dot{h}^{2}}}{\dot{h}h^{n+2}}+\frac{\sin^{n}(\frac{\Omega}{2})}{h^{n+2}}-\epsilon_n\frac{P_{1}}{h}-\sum'_{i=0}\frac{P_{(n-2i)}}{h^{n-2i}}\Big]
\nonumber\\
&+\int_{\delta}^{z_{m}}\frac{dz}{z^{d-n-2}}\Big[\frac{\sin^{n}(\frac{\Omega}{2})}{n+1}(\frac{1}{h_{c}^{n+1}}-\frac{1}{h_{0}^{n+1}})+\epsilon_n P_{1}\log(\frac{h_{c}}{h_{0}})
\nonumber\\
&-\sum'_{i=0}\frac{P_{(n-2i)}}{n-2i-1}(\frac{1}{h_{c}^{n-2i-1}}-\frac{1}{h_{0}^{n-2i-1}})\Big]
\nonumber\\
&=I_{1}+I_{2}
\end{align}
where $I_{1}$ and $I_{2}$ represent the first and second integrals, respectively. Note, in the above expression, we extract out the $\frac{1}{h}$ term in the expansion for odd $n$. This term that leads to double log term and appears only in even dimension since $n-i$ must be $1$ and $i$ is even so $n$ is odd and $d=n+3$ even. Prime on the summation means exclusion of $i=n-1$. With these explanations we have:
\begin{align}
\frac{dI_{1}}{d\delta}&=
-\frac{1}{\delta^{d-n-2}} \int_{h_{0}}^{0}dh
\Big[\frac{\sin^{n}(\theta)\sqrt{1+h^{2}+\dot{h}^{2}}}{\dot{h}h^{n+2}}+\frac{\sin^{n}(\frac{\Omega}{2})}{h^{n+2}}-\epsilon_n\frac{P_{1}}{h}-\sum'_{i=0}P_{(n-2i)}\frac{1}{h^{n-2i}}\Big]
\nonumber\\
&\equiv-\frac{1}{\delta^{d-n-2}} \int_{h_{0}}^{0}dh J_{1}(h).
\end{align}
where we have defined $J_1(h)$ as the integrand of the first line. Similarly, we derive
\begin{align}
&\frac{dI_{2}}{d\delta}=
-\frac{\sin^{n}(\frac{\Omega}{2})}{n+1}\frac{H^{n+1}}{\delta^{d-1
}}+\sum'_{i=0}\frac{P_{(n-2i)}}{n-2i-1}\frac{H^{n-2i-1}}{\delta^{d-2i-3
}}-\epsilon_n\frac{P_{1}}{\delta^{d-n-2}}\log(\frac{\delta}{H}).
\end{align}
Then, with substituting $n=d-3$ in these expression, we find:
\begin{align}
\frac{dI}{d\delta}&=\frac{dI_{1}}{d\delta}+\frac{dI_{2}}{d\delta}\nonumber\\
&= -\frac{1}{\delta} \int_{h_{0}}^{0}dh  J_{1}(h)
-\frac{\sin^{d-3}(\frac{\Omega}{2})}{d-2}\frac{H^{n+1}}{\delta^{d-1
}}+\sum'_{i=0}\frac{P_{(d-2i-3)}}{d-i-4}\frac{H^{d-2i-4}}{\delta^{d-2i-3
}}-\epsilon_d\frac{P_{1}}{\delta}\log(\frac{\delta}{H})
\end{align}
As we see, the logarithmic and power law divergences always appear in each dimensions. But, the double logarithmic term appears only in even dimensions. These terms reflect the singularity characteristic of entangling surface and denotes the geometric data contribution to entanglement entropy. Using in which $h_{c}(\delta)=\delta/H$ and $z_m=H h_0$, the entanglement entropy derived as 
\begin{align}
S_{EE}=&\frac{L^{d-1}\Omega^{d-3}}{2G_{N}}\left[q_d\log(\frac{\delta}{H h_0}) -\epsilon_dP_{1}(\log\frac{\delta}{H h_0})^2  + \cdots\right]
\end{align}
where $q_{d}$ is
\begin{align}
q_{d}= - \int_{h_{0}}^{0}dh  J_{1}(h)
\end{align}
So there are logarithmic and double logarithmic divergences for EE.


To be more specific, we consider $d=4, 5, 6$ in the following. 
For special case, $d=4$, we have a cone $c_1$ and find,
\begin{align}\label{eq 2}
  &\frac{\sin(\theta)\sqrt{1+h^{2}+\dot{h}^{2}}}{\dot{h}h^{3}}\sim -\frac{\sin(\frac{\Omega}{2})}{h^{3}}+\frac{\cos(\frac{\Omega}{2})\cot (\frac{\Omega}{2})}{8h}+\cdots .
\end{align}
Note that the next term in the above expansion includes a logarithm, but it is less singular and does not contribute to the entropy. 
Hence, by using the relations \eqref{eq 2} we can isolate the divergent part of the entropy functional and make it finite. In $d=4$ we have
\begin{align}
S_{EE}=\frac{L^{3}\Omega}{2G_{N}}(I_{1}+I_{2})
\end{align}
where
\begin{align}
I_{1}&=\int_{\delta}^{z_{m}}\frac{dz}{z} \int_{h_{0}}^{h_{c}}dh
\Big[\sin(\theta)\frac{\sqrt{1+h^{2}+\dot{h}^{2}}}{\dot{h}h^{3}}+\frac{\sin(\frac{\Omega}{2})}{h^{3}}-\frac{\cos(\frac{\Omega}{2})\cot (\frac{\Omega}{2})}{8h}\Big] \\
I_{2}&=-\int_{\delta}^{z_{m}}\frac{dz}{z} \int_{h_{0}}^{h_{c}}dh
\Big(\frac{\sin(\frac{\Omega}{2})}{h^{3}}-\frac{\cos(\frac{\Omega}{2})\cot (\frac{\Omega}{2})}{8h}\Big) 
\end{align}
Now, we differentiate $I_1$ and $I_2$ with respect to $UV$ cut-off $\delta$ and look for various divergent terms. we find
\begin{align}
\frac{dI_{1}}{d\delta}&=-\frac{1}{\delta}\int_{h_{0}}^{0}dh\Big[\sin(\theta)\frac{\sqrt{1+h^{2}+\dot{h}^{2}}}{\dot{h}h^{3}}+\frac{\sin(\frac{\Omega}{2})}{h^{3}}-\frac{\cos(\frac{\Omega}{2})\cot (\frac{\Omega}{2})}{8h}\Big]+\cdots \,,
\\
\frac{dI_{2}}{d\delta}&= \frac{-\sin(\frac{\Omega}{2})}{2}\frac{H^{2}}{\delta^{3}}  -\frac{\cos(\frac{\Omega}{2})\cot (\frac{\Omega}{2})}{8}\log 
(\frac{\delta}{H})\frac{1}{\delta}+(\frac{\sin(\frac{\Omega}{2})}{h^{2}_{0}}+\frac{\cos(\frac{\Omega}{2})\cot (\frac{\Omega}{2})}{8}\log(h_{0}))\frac{1}{\delta}+\cdots,
\end{align}
Then we find that
\begin{align}
&S_{EE}=\frac{L^{3}2\pi}{2G_{N}}\Big[\frac{H^{2}\sin(\frac{\Omega}{2})}{4\delta^{2}}-\frac{1}{16}\cos(\frac{\Omega}{2})\cot (\frac{\Omega}{2})(\log \frac{\delta}{H})^{2}+q_{4}\log (\frac{\delta}{H})+\cdots\Big],
\end{align}
where
\begin{align}
q_{4}=\frac{\sin(\frac{\Omega}{2})}{2h_{0}^{2}}+\frac{1}{8}\cos(\frac{\Omega}{2})\cot (\frac{\Omega}{2})-\int_{h_{0}}^{0}dh\Big[\frac{\sin(\theta)\sqrt{1+h^{2}+\dot{h}^{2}}}{\dot{h}h^{3}}+\frac{\sin(\frac{\Omega}{2})}{h^{3}}-\frac{\cos(\frac{\Omega}{2})\cot (\frac{\Omega}{2})}{8h}\Big]
\end{align}
As we see, there is a double logarithmic divergence for EE of a cone $c_{1}$ in $d=4$  \cite{Ref32}.

Similarly, one can calculate the entanglement entropy for cone $c_{n}$ in other dimensions. For $d=5$, one finds \begin{align}\label{eq 2-d5}
  &\frac{\sin^{2}(\theta)\sqrt{1+h^{2}+\dot{h}^{2}}}{\dot{h}h^{4}}\sim -\frac{\sin^{2}(\frac{\Omega}{2})}{h^{4}}+P_{2}\frac{1}{h^{2}}+\cdots ,
\end{align}
with $P_{2}=(4/9)\cos^{2}(\Omega/2)$. Then we have
\begin{align}
S_{EE}=\frac{L^{4}\Omega_{2}}{2G_{N}}(I_{1}+I_{2})
\end{align}
where
\begin{align}
I_{1}&=\int_{\delta}^{z_{m}}\frac{dz}{z} \int_{h_{0}}^{h_{c}}dh
\Big[\sin^{2}(\theta)\frac{\sqrt{1+h^{2}+\dot{h}^{2}}}{\dot{h}h^{4}}+\frac{\sin^{2}(\frac{\Omega}{2})}{h^{4}}-P_{2}\frac{1}{h^{2}}\Big]\,,
\\
I_{2}&=\int_{\delta}^{z_{m}}\frac{dz}{z} \int_{h_{0}}^{h_{c}}dh
\Big(-\frac{\sin^{2}(\frac{\Omega}{2})}{h^{4}}+P_{2}\frac{1}{h^{2}}\Big) 
\end{align}
Now, we differentiate these terms with respect to $UV$ cut-off $\delta$,
\begin{align}
\frac{dI_{1}}{d\delta}&=-\frac{1}{\delta}\int_{h_{0}}^{0}dh\Big[\sin^{2}(\theta)\frac{\sqrt{1+h^{2}+\dot{h}^{2}}}{\dot{h}h^{4}}+\frac{\sin^{2}(\frac{\Omega}{2})}{h^{4}}-P_{2}\frac{1}{h^{2}}\Big]+\cdots \,,\\
\frac{dI_{2}}{d\delta}&=\frac{-\sin^{2}(\frac{\Omega}{2})}{3}\frac{H^{3}}{\delta^{4}} + (\frac{\sin^{2}(\frac{\Omega}{2})}{3h_{0}}-P_{2}H)+\cdots,  
\end{align}
in which $h_{c}(\delta)=\delta/H$. Then we find that
\begin{align}
&S_{EE}=\frac{L^{4}\Omega_{2}}{2G_{N}}\Big[\frac{H^{3}\sin^{2}(\frac{\Omega}{2})}{12\delta^{3}}+q_{5}\log (\frac{\delta}{H})+\cdots\Big],
\end{align}
where
\begin{align}
q_{5}=\frac{\sin^{2}(\frac{\Omega}{2})}{3h_{0}^{3}}-P_{2}H-\int_{h_{0}}^{0}dh\Big[\sin^{2}(\theta)\frac{\sqrt{1+h^{2}+\dot{h}^{2}}}{\dot{h}h^{4}}+\frac{\sin^{2}(\frac{\Omega}{2})}{h^{4}}\Big]
\end{align}
In this case, there is only a logarithmic divergence for EE of a cone $c_{2}$ in $d=5$  \cite{Ref32}.

 For cone $c_{3}$, $d=6$, we have
\begin{align}\label{eq 2-d6}
  &\frac{\sin^{3}(\theta)\sqrt{1+h^{2}+\dot{h}^{2}}}{\dot{h}h^{5}}\sim -\frac{\sin^{3}(\frac{\Omega}{2})}{h^{5}}+P_{3}\frac{1}{h^{3}}+P_{1}\frac{1}{h}\cdots ,
\end{align}
where $P_{3}=27\cos^{2}(\frac{\Omega}{2})\sin(\frac{\Omega}{2})/(32h^{3})$ and $P_{1}=-9\cos(\frac{\Omega}{2})\cot (\frac{\Omega}{2})(31-\cos\Omega)/(4096h)$.

 Then we find
\begin{align}
S_{EE}=\frac{L^{4}\Omega_{3}}{2G_{N}}(I_{1}+I_{2})
\end{align}
where
\begin{align}
I_{1}&=\int_{\delta}^{z_{m}}\frac{dz}{z} \int_{h_{0}}^{h_{c}}dh
\Big[\sin^{3}(\theta)\frac{\sqrt{1+h^{2}+\dot{h}^{2}}}{\dot{h}h^{5}}+\frac{\sin^{3}(\frac{\Omega}{2})}{h^{5}}-P_{3}\frac{1}{h^{3}}-P_{1}\frac{1}{h}\Big]  \\
I_{2}&=\int_{\delta}^{z_{m}}\frac{dz}{z} \int_{h_{0}}^{h_{c}}dh
\Big(-\frac{\sin^{3}(\frac{\Omega}{2})}{h^{5}}+P_{3}\frac{1}{h^{3}}+P_{1}\frac{1}{h}\Big) 
\end{align}
Now, we differentiate these terms with respect to $UV$ cut-off $\delta$,
\begin{align}
\frac{dI_{1}}{d\delta}&=-\frac{1}{\delta}\int_{h_{0}}^{0}dh\Big[\sin^{3}(\theta)\frac{\sqrt{1+h^{2}+\dot{h}^{2}}}{\dot{h}h^{5}}+\frac{\sin^{3}(\frac{\Omega}{2})}{h^{5}}-P_{3}\frac{1}{h^{3}}-P_{1}\frac{1}{h}\Big]+\cdots \,,\\
\frac{dI_{2}}{d\delta}&= \frac{-\sin^{3}(\frac{\Omega}{2})}{4}\frac{H^{4}}{\delta^{5}}+P_{3}\frac{H^{2}}{2\delta^{3}}-\frac{P_{1}}{2\delta}(\log \frac{\delta}{H}) + \Big(P_{1}\log h_{0}-\frac{\sin^{3}(\frac{\Omega}{2})}{4h_{0}^{4}}-\frac{P_{3}}{2h_{0}^{2}} \Big)\frac{1}{\delta}+\cdots,
  \end{align}
then
\begin{align}
S_{EE}&=\frac{L^{4}\Omega_{3}}{2G_{N}}\Big[\frac{\sin^{3}(\frac{\Omega}{2})H^{4}}{16\delta^{4}}-P_{3}\frac{H^{2}}{4\delta^{2}}-\frac{P_{1}}{4}(\log \frac{\delta}{H})^{2} \nonumber\\
&+\Big(q_{6}+P_{1}\log h_{0}-\frac{\sin^{3}(\frac{\Omega}{2})}{4h_{0}^{4}}-\frac{P_{3}}{2h_{0}^{2}}\Big)\log (\frac{\delta}{H})+\cdots\Big],
\end{align}
where
\begin{align}
q_{6}=-\int_{h_{0}}^{0}dh\Big[\sin^{3}(\theta)\frac{\sqrt{1+h^{2}+\dot{h}^{2}}}{\dot{h}h^{5}}+\frac{\sin^{3}(\frac{\Omega}{2})}{h^{5}}-P_{3}\frac{1}{h^{3}}-P_{1}\frac{1}{h}\Big]
\end{align}
So, in $c_n$ geometries, there are logarithmic and double logarithmic divergences for EE of odd and even dimensions, respectively \cite{Ref32}.

   
\subsection{Crease $k\times R^{m}$}
\label{sec22}

The crease region $k\times R^{m}$ in time slice $t_{E}= 0 $ is given by $0\leq \rho\leq \infty,\frac{-\Omega}{2}\leq \theta \leq \frac{\Omega}{2}$, and $x^{i}\in[-\infty,\infty]$.  Recalling the metric \eqref{si} with setting $n=0$, and pick $\sigma=(\rho,\theta,x^{i},z)$ as the induced coordinates over the minimal surface. We parametrize the minimal bulk surface as $\rho=\rho(z,\theta)$. Hence the induced metric on it becomes
 \begin{equation}
   ds^{2}=\frac{l^{2}}{z^{2}}(\rho'^{2} dz^{2}+2\rho'\dot{\rho}dzd\theta+(\dot{\rho}^{2}+\rho^{2})d\theta^{2}+\sum_{i=1}^{m}(dx^{i})^{2})
 \end{equation}
 where $\dot{\rho}= \partial_{\theta}\rho$, $\rho'=\partial_{z}\rho$. Now the EE is given by
 \begin{equation}\label{eq 3}
   S_{EE}=\frac{1}{4G_{N}}\int d\sigma\sqrt{\gamma}=\frac{L^{d-1}\tilde{H}^{d-3}}{4G_{N}}\int_{\delta}^{z_{m}}dz\int_{\frac{-\Omega}{2}+\epsilon}^{\frac{\Omega}{2}-\epsilon}
   d\theta\frac{1}{z^{d-1}}\sqrt{(\rho'^{2}+1)\rho^{2}+\dot{\rho}^{2}}
 \end{equation}
 where $\delta$ is the UV cut-off, $\epsilon$ is angular cut-off, and $z_{m}$ is defined such that $\rho(Z_{m},0)=H$. We regulate $(d-3)$- dimensional space such that $x^{i}\in[-\frac{\tilde{H}}{2},\frac{\tilde{H}}{2}]$  and take $\rho=z/h(\theta)$
  with $h(\theta)$ is defined such that $h(\theta)\rightarrow0$ as $\theta\rightarrow\pm\frac{\Omega}{2}$, $\dot{h}=\partial_{\theta}h$, and $\dot{h_{0}}=h(0)=0$. Using this in \eqref{eq 3} we reach to the following functional entropy
     \begin{equation}\label{EEkRm}
    S_{EE}=\frac{L^{d-1}\tilde{H}^{d-3}}{2G_{N}}\int_{\delta}^{z_{m}}\frac{dz}{z^{d-2}}\int_{h_{0}}^{h_{c}} \frac{dh}{\dot{h}}\frac{\sqrt{1+h^{2}+\dot{h}^{2}}}{h^{2}}
     \end{equation}
 From the variation of entropy functional, we find the equation of motion for $h$
\begin{equation}
  h(1+h^{2})\ddot{h}+(d-1)\dot{h}^{2}+h^{4}+dh^{2}+d-1=0
\end{equation}
The conserved quantity corresponds to the above equation defined as \cite{Ref32}
\begin{equation}\label{Kd}
 K_{d} = \frac{(1+h^{2})^{\frac{d-1}{2}}}{h^{(d-1)}\sqrt{1+h^{2}+\dot{h}^{2}}}= \frac{(1+h_{0}^{2})^{\frac{d-2}{2}}}{h_{0}^{(d-1)}}.
\end{equation}
 Due to decreasing of $h$ near the boundary, $\dot{h}$ should be negative. Hence we find that
\begin{equation}
 \dot{h} =- \frac{\sqrt{1+h^{2}}\sqrt{(1+h^{2})^{d-2}-K_{d}^{2}h^{2(d-1)}}}{K_{d}h^{d-1}}.
\end{equation}
Now we analyze the divergence of the above entropy functional. Near the boundary, the integrand of \eqref{EEkRm} in the asymptotic limit behaves as
\begin{equation}
  \frac{\sqrt{1+h^{2}+\dot{h}^{2}}}{\dot{h}h^{2}}\thicksim - \frac{1}{h^{2}} - \frac{1}{2} K_{d}^{2}h^{2(d-2)}+ ...
\end{equation}
Hence we can isolate the divergent part of integrals in the following way
\begin{align}
I&= \int_{\delta}^{z_{m}}\frac{dz}{z^{d-2}}\int_{h_{0}}^{h_{c}} \frac{dh}{\dot{h}}\frac{\sqrt{1+h^{2}+\dot{h}^{2}}}{h^{2}}
  \nonumber\\
&=\int_{\delta}^{z_{m}}\frac{dz}{z^{d-2}}\int_{h_{0}}^{h_{c}}dh
  [\frac{\sqrt{1+h^{2}+\dot{h}^{2}}}{\dot{h}h^{2}}+\frac{1}{h^{2}}]+\int_{\delta}^{z_{m}}\frac{dz}{z^{d-2}}(\frac{1}{h_{c}}
  -\frac{1}{h_{0}})
  \nonumber \\
 &= I_{1}+I_{2}
\end{align}
where $I_{1}$ and $I_{2}$ represent the first and second integrals, respectively.

First we consider $I_{1}$, and differentiate it with respect to $UV$ cut-off  $\delta$ and look for various divergent terms. we find
\begin{align}
  \frac{dI_{1}}{d\delta}&= \frac{-1}{\delta^{d-2}} \int_{h_{0}}^{h_{c}(\delta)}dh
  [\frac{\sqrt{1+h^{2}+\dot{h}^{2}}}{\dot{h}h^{2}}+\frac{1}{h^{2}}]
  \nonumber\\
 & =\frac{-1}{\delta^{d-2}}\int_{h_{0}}^{0}dh
  [\frac{\sqrt{1+h^{2}+\dot{h}^{2}}}{\dot{h}h^{2}}+\frac{1}{h^{2}}]+\frac{1}{2}\frac{K_{d}^{2}}{H^{2d-3}}\delta^{d-1}+\cdots,
\end{align}
where we have used Taylor expansion in the second line. Similarly, for $I_{2}$ we find that
 \begin{equation}
 \frac{dI_{2}}{d\delta}=-\frac{1}{\delta^{d-2} h_{c}(\delta)}+\frac{1}{\delta^{d-2} h_{0}}=-\frac{H}{\delta^{d-1}}+\frac{1}{\delta^{d-2}h_{0}}+\cdots,
 \end{equation}
 where in the above we used $h_{c}(\delta)=\delta/H$. So  we have
 \begin{equation}
 S_{EE}=\frac{L^{d-1}\tilde{H}^{d-3}}{2G_{N}}\Big[\frac{1}{(d-2)}\frac{H}{\delta^{d-2}}+\Big(\int_{h_{0}}^{0}dh
  (\frac{\sqrt{1+h^{2}+\dot{h}^{2}}}{\dot{h}h^{2}}+\frac{1}{h^{2}})- \frac{1}{h_{0}}\Big)\frac{1}{(d-3)}\frac{1}{\delta^{d-3}}+\cdots\Big].
 \end{equation}
As we see, there is no logarithmic divergence due to singularity of entangling surface, but there is a new power law divergent term of order $1/\delta^{d-3}$ which is due to singularity and does not arise in smooth entangling surface \cite{Ref32}.


\section{Holographic entanglement entropy of singular surfaces in relevant perturbed theory}
\label{sec4}
In this section, we will study entanglement entropy of the singular geometries in the form $k\times R^{m}$, $c_{n}$ and $c_{n}\times R^{m}$ for relevant perturbed conformal field theory in $d$- dimensions which is dual to Einstein gravity. In three dimensions, the kink geometry considered previously in \cite{Ref41}. Here, first we focus on $d$ dimensions, then in order to get explicit solution, we zoom in to $d=4,5,6$ dimensions.

\subsection{Crease $c_{n}\times R^{m}$  }
\label{sec3.3}
Let us start with singular surfaces in the form of $c_{n}\times R^{m}$ in asymptotic anti-de Sitter space time background.  We choose the metric ansatz in the form
\begin{equation}\label{metric-3.1}
 ds ^{2}= \frac{L^{2}}{z^{2}}\Big(-dt^{2}+d\rho^{2}+\rho^{2}(d\theta^{2}+\sin^{2}\theta  d\Omega_{n}^{2})+\frac{dz^{2}}{f(z)}+\sum_{i=1}^{m}(dx^{i})^{2}\Big),
\end{equation}
where $f(z)\rightarrow1$ as $z\rightarrow0$.  Similar to previous section and symmetry discussions, we parameterize the bulk minimal surface as $ \rho=\rho(z,\theta)$. Then the induced metric in entangling surface in time slice $t=0$ becomes
\begin{align}
 ds ^{2}&=\gamma_{ab}dx^adx^b  \nonumber\\
&=\frac{L^{2}}{z^{2}}\Big( \big(\rho^{'2}+\frac{1}{f(z)}\big)dz^{2}+2\rho'\dot{\rho}dzd\theta+(\dot{\rho}^{2}+\rho^{2})
 d\theta^{2}+\rho^{2}\sin^{2}\theta  d\Omega_{n}^{2}+\sum_{i=1}^{m}(dx^{i})^{2}),
\end{align}
so that
\begin{equation}
  \sqrt{\gamma}=\frac{L^{d-1}}{z^{d-1}}\rho^{n}\sin^{n}\theta\sqrt{\big(\rho^{'2}+\frac{1}{f(z)}\big)\rho^{2}+\frac{1}{f(z)}\dot{\rho}^{2}}.
\end{equation}
By the RT prescription, the entanglement entropy is derived as
\begin{equation}\label{eq2}
S_{EE}=\frac{1}{4G_{N}}\int dzd\theta  \sqrt{\gamma} = \frac{L^{d-1}\Omega_{n}\tilde{H}^{m}}{4G_{N}}\int dzd\theta   \frac{\rho^{n}\sin^{n}\theta}{z^{d-1}\sqrt{f}}\sqrt{(f\rho^{'2}+1)\rho^{2}+\dot{\rho}^{2}}
\end{equation}
where $\tilde{H}^{m}$ is volume of $m$-dimensional space. By extremizing the above action we can derive the equation of motion for $\rho(z,\theta)$, which reads
\begin{align}\label{eq-rho}
&2fz\sin\theta\rho^{2}(\rho^{2}+\dot{\rho}^{2})\rho''
+2z\sin\theta\rho^{2}(1+f\rho^{'2})\ddot{\rho}
-4fz\sin\theta\rho^{2} \rho' \dot{\rho}\dot{\rho}'
+z\sin\theta\rho^{2}\rho'(\rho^{2}+\dot{\rho}^{2})f'
\nonumber\\
&-2z\sin\theta\rho((n+1)(1+f\rho^{'2})\rho^{2}+(n+2)\dot{\rho}^{2})
-2(d-1)\sin\theta\rho^{2}f\rho'((1+f\rho^{'2})\rho^{2}+\dot{\rho}^{2})
\nonumber\\
&+2nz\cos\theta\dot{\rho}((1+f\rho^{'2})\rho^{2}+\dot{\rho}^{2})=0
\end{align}
where $\ddot{\rho}=\partial_{\theta}^{2}\rho$, $\rho''=\partial_{z}^{2}\rho$ and ${\dot{\rho}'}=\partial_{\theta}\partial_{z}\rho$.
For small deformation we make the following ansatz:
\begin{equation}\label{eqans}
  \rho(z,\theta)=\rho_{0}+\delta\rho=\frac{z}{h(\theta)}+\delta f g(\theta)z,\qquad  f=1+\delta f
\end{equation}
with inserting the ansatz \eqref{eqans} in the equation of motion for $\rho(z,\theta)$ we derive the equations for $h(\theta)$ and $g(\theta)$ as
\begin{align}\label{eq-h}
&\sin(\theta) h(1+h^{2})\ddot{h}+\sin(\theta)(nh^{2}+d-1)\dot{h}^{2}
\nonumber\\
&+n\cos(\theta)h\dot{h}(1+h^{2}+\dot{h}^{2})+\sin(\theta)(1+h^{2})(d-1+(n+1)h^{2})=0,
\end{align}
and
\begin{align}\label{eq4}
&2\sin(\theta)h^{3}(1+h^{2})\delta f \ddot{g}
\nonumber\\
&+2h^{2}\Big[2\sin(\theta)\dot{h}z\delta f'+\Big(2\dot{h}(d+1+(n+2)h^{2})\sin(\theta)+
nh(1+h^{2}+3\dot{h}^{2})\cos(\theta)\Big)\delta f\Big]\dot{g}
\nonumber\\
&+2h\Big[\sin(\theta)(h^{2}+\dot{h}^{2})(\delta f''z^{2}+2z{\delta f'})+2\sin(\theta)(z\delta{ f'}+(2+h^{2})\delta f)(-h\ddot{h}+2\dot{h}^{2})
\nonumber\\
&-\sin(\theta)(z{\delta f'}+3\delta f)\Big((d+1)\dot{h}^{2}+(d-1)(h^{2}+1)\Big)
\nonumber\\
&- 2(\delta f(2+h^{2})+z\delta f')\Big((d-1)\sin(\theta)+n\cos(\theta)h\dot{h}\Big)
\nonumber\\
&-\sin(\theta)h^{2}\Big(2(n+1)z\delta f'+((n+2)\dot{h}^{2}+(n+1)(3h^{2}+5))\delta f\Big)\Big]g
\nonumber\\
&+\sin(\theta)\Big[(\dot{h^{2}}+h^{2})z\delta f'-2(\ddot{h}h+(d-1)\dot{h}^{2}+(d+n)h^{2}+2(d-1))\delta f\Big]
\nonumber\\
&-2n\cos(\theta)h\dot{h}\delta f=0.
\end{align}
We insert $\delta f(z)=\mu^{2\alpha} z^{2\alpha}$  in  Eq.\eqref{eq4} and by using the equation of motion for $h$ we can reach to the following equation of motion for $g(\theta)$  for various $\alpha$ :
\begin{align}
&\sin(\theta)h^{3}(1+h^{2})^{2}\ddot{g}
\nonumber\\
&+h^{2}(1+h^{2})\Big[2(2\alpha+d+1+(n+2)h^{2})\dot{h}\sin(\theta)+nh(1+h^{2}+3\dot{h}^{2})\cos(\theta)\Big]\dot{g}
\nonumber\\
&+h\Big[\sin(\theta)(1+h^{2})(h^{2}+\frac{1}{3} \dot{h}^{2})2\alpha(2\alpha+1)
+2\sin(\theta)(2\alpha+2+h^{2})\Big((nh^{2}+d-1)\dot{h}^{2}
\nonumber\\
&+n\cot(\theta)h\dot{h}(1+h^{2}+\dot{h}^{2})+(1+h^{2})(d-1+(n+1)h^{2})\Big)
\nonumber\\
&-\sin(\theta)(1+h^{2})(2\alpha+3)\Big(\frac{1}{3}(3d-4\alpha-5)\dot{h}^{2}+(d-1)(h^{2}+1)\Big)
\nonumber\\
&-2(1+h^{2})(2+2\alpha+h^{2})\Big((d-1)\sin(\theta)+ n\cos(\theta)h\dot{h}\Big)
\nonumber\\
&-\sin(\theta)h^{2}(1+h^{2})\Big((n-2)\dot{h}^{2}+(n+1)(3h^{2}+4\alpha+5)\Big)\Big]g
\nonumber\\
&+\Big[\sin (\theta ) \left(\dot{h}^2 \left(h^2 (\alpha-d+n+1)+\alpha\right)+h^2 \left((\alpha-d+1) h^2+\alpha-2 d+2\right)-d+1\right) \nonumber\\
&+n\cos (\theta )h \dot{h}^3 \Big]=0,    
\end{align}
 Plugging the small deformation $\rho(z,\theta)=\rho_{0}+\delta\rho$ in the entropy functional \eqref{eq2} and use the Taylor expansion we get, up to first order, the following perturbed entropy functional:
\begin{align}
&S_{EE}=\frac{L^{d-1}}{4G_{N}}\Omega_{n}\tilde{H}^{m}\int_{\delta}^{z_{m}}dz \int_{\frac{-\Omega}{2}+\epsilon}^{\frac{\Omega}{2}-\epsilon}d\theta
\Big[\frac{\sin^{n}(\theta)\rho_{0}^{n}R_0}{z^{d-1}}
-\frac{\rho_{0}^{n}\sin^{n}(\theta)(\rho_{0}^{2}+\dot{\rho_{0}}^{2}) \delta f}{2z^{d-1}R_0}
\nonumber\\
&\qquad +\rho_{0}^{n-1}(\sin^{n}(\theta))\frac{\rho_{0}^{3}\rho_{0}'\delta\rho'+((n+1)\rho_{0}^{2}(1+\rho_{0}^{'2})+n\dot{\rho_{0}}^{2})\delta\rho+
\rho_{0}\dot{\rho_{0}}\delta\dot{\rho}}{z^{d-1}R_0}\Big]
\end{align}
where $R_0\equiv\sqrt{\rho_{0}^{2}(1+\rho_{0}^{'2})+\dot{\rho_{0}}^{2}}$ and $\delta$ is the UV cut-off, $\epsilon$ is angular cut-off, and $z_{m}$ is defined such that $\rho(z_{m},0)=H $. Using integration by parts and equation of motion for $\rho_{0}$, we will find the following entropy functional
\begin{align}\label{eq5}
&S_{EE}=\frac{L^{d-1}}{4G_{N}}\Omega_{n}\tilde{H}^{m}\int_{\delta}^{z_{m}}dz \int_{\frac{-\Omega}{2}+\epsilon}^{\frac{\Omega}{2}-\epsilon}d\theta
\Big[\frac{\sin^{n}(\theta)\rho_{0}^{n}R_0}{z^{d-1}}
-\frac{\rho_{0}^{n}\sin^{n}(\theta)(\rho_{0}^{2}+\dot{\rho_{0}}^{2}) \delta f}{2z^{d-1}R_0}
\nonumber\\
&\qquad +\partial_{\theta}\Big(
\frac{\sin^{n}(\theta)\rho_{0}^{n}\dot{\rho_{0}}\delta\rho}{z^{d-1}R_0}\Big)
+\partial_{z}\Big(
\frac{\sin^{n}(\theta)\rho_{0}^{n+2}\rho_{0}'\delta\rho}
{z^{d-1}R_0}\Big)\Bigg].
\end{align}
Now by substituting the ansatz \eqref{eqans} in \eqref{eq5}, we reach to the following entropy functional,
\begin{equation}
S_{EE}=\frac{L^{d-1}\Omega_{n}\tilde{H}^{m}}{2G_{N}}\Big (I_{1}+I_{2}+I_{3}+I_{4}\Big),
\end{equation}
where $I_{1}$, $I_{2}$, $I_{3}$, and $I_{4}$ defined as:
\begin{align}
I_{1}&=\int_{\delta}^{z_{m}}dz \int_{0}^{\frac{\Omega}{2}-\epsilon}d\theta
\Big[\frac{\sin^{n}(\theta)\rho_{0}^{n}R_0}{z^{d-1}}\Big]
\nonumber\\
 &=\int_{\delta}^{z_{m}}\frac{dz}{z^{d-n-2}} \int_{h_{0}}^{h_{c}}dh
\frac{\sin^{n}(\theta)\sqrt{1+h^{2}+\dot{h}^{2}}}{\dot{h}h^{n+2}},
\\
I_{2}&=\int_{\delta}^{z_{m}}dz \int_{0}^{\frac{\Omega}{2}-\epsilon}d\theta
\frac{-\rho_{0}^{n}\sin^{n}(\theta)(\rho_{0}^{2}+\dot{\rho_{0}}^{2}) \delta f}{2z^{d-1}R_0}
\nonumber\\
&=\int_{\delta}^{z_{m}}dz\frac{\delta f}{2z^{d-n-2}} \int_{h_{0}}^{h_{c}}dh
\sin^{n}(\theta)\frac{-(h^{2}+\dot{h}^{2}) }{\dot{h}h^{n+2}\sqrt{1+h^{2}+\dot{h}^{2}}},
\\
I_{3}&=\int_{\delta}^{z_{m}}dz \int_{0}^{\frac{\Omega}{2}-\epsilon}d\theta
\partial_{\theta}\Big(
\frac{\sin^{n}(\theta)\rho_{0}^{n}\dot{\rho_{0}}\delta\rho}{z^{d-1}R_0}\Big)
\nonumber\\
&=\int_{\delta}^{z_{m}}dz\frac{\delta f}{z^{d-n-2}}\frac{-\sin^{n}(\theta)\dot{h}g(\theta)}{h^{n}\sqrt{1+h^{2}+\dot{h}^{2}}}|_{h=h_{c}(\delta)},
\\
I_{4}&=\int_{\delta}^{z_{m}}dz \int_{0}^{\frac{\Omega}{2}-\epsilon}d\theta \partial_{z}\Big(
\frac{\sin^{n}(\theta)\rho_{0}^{n+2}\acute{\rho_{0}}\delta\rho}
{z^{d-1}R_0}\Big)
\nonumber\\
&=\int_{\delta}^{z_{m}}dz\frac{z\delta f'-(d-n-3)\delta f}{z^{d-n-2}} \int_{h_{0}}^{h_{c}}dh
\sin^{n}(\theta)\frac{g}{\dot{h}h^{n+1}\sqrt{1+h^{2}+\dot{h}^{2}}},
\end{align}
where we have changed the integration variable to $h(\theta)$. We have also defined
$h_{0}=h(0)$ and $h_{c}=h(\frac{\Omega}{2}-\epsilon)$ and used $\dot{h}_{0}(0)=0$ in getting the boundary terms.

To extract the logarithmic divergence, we must find the asymptotic behavior of integrand in terms of $h$, where  $h\rightarrow0$. Similar to the previous section, consider $y=\sin(\theta)$ and change the independent variable  from $\theta$ to $h$, and find the equation of motion of $y=y(h)$. Using the relations 
\begin{align}
&\dot{h}=\frac{\sqrt{1-y^{2}}}{\dot{y}(h)}   \nonumber\\
&\ddot{h}=-\frac{y\dot{y}^{2}+(1-y^{2})\ddot{y}}{\dot{y}^{3}}  \nonumber\\
&\ddot{g}(\theta)=\frac{\sqrt{1-y^{2}}}{\dot{y}(h)} \frac{d}{dh}\Big(\dot{g}(h)\frac{\sqrt{1-y^{2}}}{\dot{y}(h)}\Big)
\nonumber\\
&\dot{g}(\theta)=\dot{g}(h)\frac{\sqrt{1-y^{2}}}{\dot{y}(h)}
\end{align}
where $\dot{y}(h)=\frac{dy}{dh}$. Finally, we reach to the following equations
\begin{align}\label{y-equation}
&h(1+h^{2})y(1-y^{2})\ddot{y}-(1+h^{2})(d-1+(n+1)h^{2})y\dot{y}^{3}+(1+h^{2})h((1+n)y^{2}-n)\dot{y}^{2}
\nonumber\\
&-nh(1-y^{2})^{2}-(nh^{2}+d-1)y(1-y^{2})\dot{y}=0
\end{align}
and
\begin{align}\label{g-equation}
&h^{3}(1+h^{2})^{2}(1-y^{2})y\dot{y}\ddot{g}
+h^{2}(1+h^{2})\Big[2nh(1-y^{2})^{2}+(3+d+4\alpha+(4+n)h^{2})(1-y^{2})y\dot{y}
\nonumber\\
&-(1+h^{2})(d-1+(1+n)h^{2})y\dot{y}^{3}\Big]\dot{g}
+h\Big[2nh(2\alpha+2+h^{2})(1-y^{2})^{2}
\nonumber\\
&+\Big((1+2\alpha)(1+d+2\alpha)+(5+3n+8\alpha+4\alpha(n+\alpha)-d(1+2\alpha))h^{2}+(2+n)h^{4}\Big)y(1-y^{2})\dot{y}
\nonumber\\
&-(1+h^{2})\Big((d-1)(3+2\alpha)+(n+d(3+2\alpha)-2(1+2\alpha(1+\alpha)))h^{2}+(1+n)h^{4}\Big)y\dot{y}^{3}\Big]g
\nonumber\\
&+\Big[nh(1-y^{2})^{2}+(\alpha+(\alpha+n+1-d)h^{2})y(1-y^{2})\dot{y}+(1+h^{2})(1-d+(1-d+\alpha)h^{2})y\dot{y}^{3}\Big]=0
\end{align}
In Appendix \ref{A}, we solved equations \eqref{y-equation} and \eqref{g-equation} perturbatively near the boundary with initial conditions  $y=\sin(\Omega/2)$ at $h=0$, and for $g$, we adopt a solution such that $\rho$ becomes finite in the limit $h\rightarrow 0$ and $\delta\rightarrow 0$. Series solutions for $y$ and $g$ are different for odd and even dimensions. From \eqref{Y} and \eqref{y-even-d} we can write $y$ series as
\begin{align}
y(h)&=\sum_{i=0}a_{2i}h^{2i}+\epsilon_d\Big[\log h\sum_{i=k}\tilde{a}_{2i}h^{2i}+(\log h)^2\sum_{i=k}\hat{a}_{2i}h^{2i}+\cdots\Big].
\end{align}
where $\epsilon_d$ was introduced in \eqref{epsilon}, so that logarithmic terms appear only in even dimensions. Similarly, the series of $g$ can be written as, 
\begin{align}\label{g}
g(h)&=h^{d-2\alpha-2}\Big[\sum_{i=0}c_{2i}h^{2i-1}+\epsilon_d\log h\sum_{i=0}\tilde{c}_{2i}h^{2i-1}+\epsilon_d(\log h)^2\sum_{i=0}\hat{c}_{2i}h^{2i-1}+\cdots\Big]\nonumber\\
&+\xi_{\alpha j} b_{2j}\log h+\sum_{i=0}b_{2i}h^{2i-1}+\epsilon_d\log h\sum_{i=0}\tilde{b}_{2i}h^{2i-1}+\epsilon_d(\log h)^2\sum_{i=0}\hat{b}_{2i}h^{2i-1}+\cdots.
\end{align}
where $\xi_{\alpha j}=\delta_{2\alpha,d-2j-2}$ with $j$ a positive integer. A few coefficients of the above expansion are found in Appendix \ref{A}. 

Using the ansatz \eqref{eqans} at $\rho=H$, $z=\delta$, and $h_{c}=h(\frac{\Omega}{2}-\epsilon)$, we find that $UV$ cut-off expansion of $h_{c}$ becomes as
\begin{align}\label{hc}
h_{c}(\delta)&=\frac{\delta}{H}+(\mu H)^{2\alpha}\Big(\frac{\delta}{H}\Big)^{2\alpha+2}g(\frac{\delta}{H})
\end{align}


In the following we analyze the divergence of entropy functional. Near the boundary, integrands of $I_{1}$, $I_{2}$, $I_{3}$, and $I_{4}$ in the asymptotic limit behave as
\begin{align}\label{series1}
\frac{\sin^{n}(\theta)\sqrt{1+h^{2}+\dot{h}^{2}}}{\dot{h}h^{n+2}}&\sim \sum_{i=0}\Big(P_{n-2i+2}+\epsilon_d\tilde{P}_{n-2i+2}\log h+\epsilon_d\hat{P}_{n-2i+2}\log^2h+\cdots\Big)\frac{1}{h^{n-2i+2}}\\
\frac{-\sin^{n}(\theta)(h^{2}+\dot{h}^{2})}{\dot{h}h^{n+2}\sqrt{1+h^{2}+\dot{h}^{2}}}&\sim
\sum_{i=0}\Big(Q_{n-2i+2}+\epsilon_d\tilde{Q}_{n-2i+2}\log h+\epsilon_d\hat{Q}_{n-2i+2}\log^2h+\cdots\Big)\frac{1}{h^{n-2i+2}}\\
\frac{-\sin^{n}(\theta)\dot{h}g}{h^{n}\sqrt{1+h^{2}+\dot{h}^{2}}}&\sim 
\sum_{i=0}\Big(M_{n-2i}+\epsilon_d\tilde{M}_{n-2i}\log h+\epsilon_d\hat{M}_{n-2i}\log^2h+\cdots\Big)\frac{g(h)}{h^{n-2i}}\\
\frac{\sin^{n}(\theta)g}{\dot{h}h^{n+1}\sqrt{1+h^{2}+\dot{h}^{2}}}&\sim 
\sum_{i=0}\Big(N_{n-2i-1}+\epsilon_d\tilde{N}_{n-2i-1}\log h+\epsilon_d\hat{N}_{n-2i-1}\log^2h+\cdots\Big)\frac{g(h)}{h^{n-2i-1}},
\end{align}
again log terms only contribute to even dimensions. Coefficients $P$'s, $Q$'s, $M$'s, and $N$'s are found in the Appendix \ref{B}.

Let us firstly consider the general dimensions, then we study special cases $d=5,6$.

\subsubsection{General case:}

In the following, we want to know  under what conditions the $log$ or the powers of $log$ terms appear in the $UV$-expansion of the entanglement entropy. Let us start with $I_{1}$
\begin{align}
&I_{1} =\int_{\delta}^{z_{m}}\frac{dz}{z^{d-n-2}} \int_{h_{0}}^{h_{c}}dh
\frac{\sin^{n}(\theta)\sqrt{1+h^{2}+\dot{h}^{2}}}{\dot{h}h^{n+2}}
\end{align}
where $d=n+m+3$. Now we use the series expansion in \eqref{series1} to isolate the divergence terms near the boundary as follows,
\begin{align}\label{I1}
I_{1}&=\int_{\delta}^{z_{m}}\frac{dz}{z^{d-n-2}} \int_{h_{0}}^{h_{c}}dh
\frac{\sin^{n}(\theta)\sqrt{1+h^{2}+\dot{h}^{2}}}{\dot{h}h^{n+2}}
\nonumber\\
&=\int_{\delta}^{z_{m}}\frac{dz}{z^{m+1}} \int_{h_{0}}^{h_{c}}dh J_1(h)
\nonumber\\
&+\int_{\delta}^{z_{m}}\frac{dz}{z^{m+1}}\Bigg[\bar{\epsilon}_{n}\Big(P_{1}\log(\frac{h_{c}}{h_{0}})+\frac{1}{2}\epsilon_d\tilde{P}_{1}\log^2(\frac{h_{c}}{h_{0}})+\cdots\Big)  \nonumber\\
&-\sum'_{i=0}\Bigg(\mathcal{P}_{(n-2i+2)}\Big(\frac{1}{h_{c}^{n-2i+1}}-\frac{1}{h_{0}^{n-2i+1}}\Big)
+\epsilon_d\widetilde{\mathcal{P}}_{(n-2i+2)}\Big(\frac{\log h_c}{h_{c}^{n-2i+1}}-\frac{\log h_0}{h_{0}^{n-2i+1}}\Big)
\nonumber\\
&+\epsilon_d\hat{\mathcal{P}}_{(n-2i+2)}\Big(\frac{\log^2 h_c}{h_{c}^{n-2i+1}}-\frac{\log^2 h_0}{h_{0}^{n-2i+1}}\Big)
+\cdots\Bigg)\Bigg].
\end{align}
in which
\begin{align}\label{J1}
J_1(h)&=\sin^{n}(\theta)\frac{\sqrt{1+h^{2}+\dot{h}^{2}}}{\dot{h}h^{n+2}}-\bar{\epsilon}_{n}\frac{P_1}{h}-\sum'_{i=0}\Big(P_{n-2i+2}+\epsilon_d\tilde{P}_{n-2i+2}\log h+\cdots\Big)\frac{1}{h^{n-2i+2}}\nonumber\\
\mathcal{P}_{k}&\equiv -\frac{1}{(k-1)}P_k-\frac{\epsilon_d}{(k-1)^2}\tilde{P}_k-\frac{2\epsilon_d}{(k-1)^3}\hat{P}_k+\cdots\nonumber\\
\tilde{\mathcal{P}}_{k}&\equiv -\frac{1}{(k-1)}\tilde{P}_k-\frac{2}{(k-1)^2}\hat{P}_k+\cdots\nonumber\\
\hat{\mathcal{P}}_{k}&\equiv -\frac{1}{(k-1)}\hat{P}_k+\cdots
\end{align}
Similarly for $I_2$, $I_3$ and $I_4$:
\begin{align}
{I_{2}}&=\int_{\delta}^{z_{m}}dz\frac{\delta f}{2z^{d-n-2}} \int_{h_{0}}^{h_{c}}dh
\frac{-\sin^{n}(\theta)(h^{2}+\dot{h}^{2})}{\dot{h}h^{n+2}\sqrt{1+h^{2}+\dot{h}^{2}}}
\nonumber\\
&=\int_{\delta}^{z_{m}}dz\frac{\delta f}{2z^{m+1}}\int_{h_{0}}^{h_{c}}dhJ_2(h)
\nonumber\\
&+\int_{\delta}^{z_{m}}dz\frac{\delta f}{2z^{m+1}}\Bigg[\bar{\epsilon}_{n}\Big(Q_{1}\log(\frac{h_{c}}{h_{0}})+\frac{1}{2}\epsilon_d\tilde{Q}_{1}\log^2(\frac{h_{c}}{h_{0}})+\cdots\Big)  \nonumber\\
&-\sum'_{i=0}\Bigg(\mathcal{Q}_{(n-2i+2)}\Big(\frac{1}{h_{c}^{n-2i+1}}-\frac{1}{h_{0}^{n-2i+1}}\Big)
+\epsilon_d\tilde{\mathcal{Q}}_{(n-2i+2)}\Big(\frac{\log h_c}{h_{c}^{n-2i+1}}-\frac{\log h_0}{h_{0}^{n-2i+1}}\Big)  \nonumber\\
&+\epsilon_d\hat{\mathcal{Q}}_{(n-2i+2)}\Big(\frac{\log^2 h_c}{h_{c}^{n-2i+1}}-\frac{\log^2 h_0}{h_{0}^{n-2i+1}}\Big)+\cdots\Bigg)\Bigg],
\end{align}
with
\begin{align}\label{J2}
J_2(h) &=\frac{-\sin^{n}(\theta)(h^{2}+\dot{h}^{2})}{\dot{h}h^{n+2}\sqrt{1+h^{2}+\dot{h}^{2}}}-\bar{\epsilon}_{n}\frac{Q_1}{h} -\sum'_{i=0}\Big(Q_{n-2i+2}+\epsilon_d\tilde{Q}_{n-2i+2}\log h+\cdots\Big)\frac{1}{h^{n-2i+2}}
\nonumber\\
\mathcal{Q}_{k}&\equiv -\frac{1}{(k-1)}Q_k-\frac{\epsilon_d}{(k-1)^2}\tilde{Q}_k-\frac{2\epsilon_d}{(k-1)^3}\hat{Q}_k+\cdots\nonumber\\
\tilde{\mathcal{Q}}_{k}&\equiv -\frac{1}{(k-1)}\tilde{Q}_k-\frac{2}{(k-1)^2}\hat{Q}_k+\cdots\nonumber\\
\hat{\mathcal{Q}}_{k}&\equiv -\frac{1}{(k-1)}\hat{Q}_k+\cdots
\end{align}
\begin{align}
I_{3}&=\int_{\delta}^{z_{m}}dz\frac{\delta f}{z^{d-n-2}}\frac{-\sin^{n}(\theta)\dot{h}g(\theta)}{h^{n}\sqrt{1+h^{2}+\dot{h}^{2}}}
\nonumber\\
&=\mu^{2\alpha}\int_{\delta}^{z_{m}}dz\frac{1}{z^{m+1-2\alpha}}\Big[\sum_{i=0}\Big(M_{n-2i}+\epsilon_d\tilde{M}_{n-2i}\log h+\epsilon_d\hat{M}_{n-2i}\log^2h+\cdots\Big)\frac{g(h)}{h^{n-2i}}\Big]
,\\
I_{4} &=\int_{\delta}^{z_{m}}dz\frac{z\delta f'-(d-n-3)\delta f}{z^{d-n-2}} \int_{h_{0}}^{h_{c}}dh
\frac{\sin^{n}(\theta)g(h)}{\dot{h}h^{n+1}\sqrt{1+h^{2}+\dot{h}^{2}}}
\nonumber\\
&=\mu^{2\alpha}\int_{\delta}^{z_{m}}\frac{2\alpha-m}{z^{m+1-2\alpha}} \int_{h_{0}}^{h_{c}}dh J_{4}(h)
\nonumber\\
&+\mu^{2\alpha}\int_{\delta}^{z_{m}}\frac{2\alpha-m}{z^{m+1-2\alpha}} \int_{h_{0}}^{h_{c}}dh  \Big[\sum_{i=0}\Big(N_{n-2i-1}+\epsilon_d\tilde{N}_{n-2i-1}\log h+\cdots\Big)\frac{g(h)}{h^{n-2i-1}}\Big]
\nonumber\\
&=\mu^{2\alpha}\int_{\delta}^{z_{m}}\frac{2\alpha-m}{z^{m+1-2\alpha}} \Big(\int_{h_{0}}^{h_{c}}dh J_{4}(h)+K_4(h_0)\Big)
\nonumber\\
&+\mu^{2\alpha}\int_{\delta}^{z_{m}}\frac{2\alpha-m}{z^{m+1-2\alpha}}  \Bigg\{
\epsilon_{n}\xi_{\alpha j} b_{2j}\Big(\frac{1}{2}N_1\log^2 (\frac{\delta}{H})+\frac{1}{3}\epsilon_d\tilde{N}_1\log^3 (\frac{\delta}{H})+\cdots\Big)
\nonumber\\
&+\bar{\epsilon}_{n}\sum_{k=0}\Big[b_{2k}N_{(2k)}\log (\frac{\delta}{H})+\frac{1}{2}\epsilon_d(b_{2k}\tilde{N}_{(2k)}+\tilde{b}_{2k}N_{(2k)})\log^2 (\frac{\delta}{H})
\nonumber\\
&+\frac{1}{3}\epsilon_d(\tilde{b}_{2k}\tilde{N}_{(2k)}+b_{2k}\hat{N}_{(2k)})\log^3 (\frac{\delta}{H})+\cdots\Big]
\nonumber\\
&+\zeta_{\alpha i}\sum_{k=0}\Big[c_{2k}N_{(2k+d-2\alpha-2)}\log (\frac{\delta}{H})+\frac{1}{2}\epsilon_d(c_{2k}\tilde{N}_{(2k+d-2\alpha-2)}+\tilde{c}_{2k}N_{(2k+d-2\alpha-2)})\log^2 (\frac{\delta}{H})
\nonumber\\
&+\frac{1}{3}\epsilon_d(\tilde{c}_{2k}\tilde{N}_{(2k+d-2\alpha-2)}+c_{2k}\hat{N}_{(2k+d-2\alpha-2)})\log^3 (\frac{\delta}{H})+\cdots\Big]
\nonumber\\
&+\xi_{\alpha j} b_{2j}\sum_{i=0}h^{2i-n+2}\Big(\mathcal{N}_{(n-2i-1)}+\tilde{\mathcal{N}}_{(n-2i-1)}\log h+\cdots\Big)
\nonumber\\
&+\sum_{i,k=0}\Big[h^{2i+2k-n+1}\Big(\mathcal{N}^{(b)}_{(n-2i-1)}+\epsilon_d\tilde{\mathcal{N}}^{(b)}_{(n-2i-1)}\log h+\cdots\Big)
\nonumber\\
&+h^{2i+2k+m+2-2\alpha}\Big(\mathcal{N}^{(c)}_{(n-2i-1)}+\epsilon_d\tilde{\mathcal{N}}^{(c)}_{(n-2i-1)}\log h+\cdots\Big)\Big]\Bigg\},
\end{align}
where $\zeta_{\alpha i}=\delta_{2\alpha,2i+2k+d-n-1}$ and $\mathcal{N}$'s are combinations of $N$'s, and $\mathcal{N}^{(b)}$'s of  $N$'s, $b$'s, $\tilde{b}$'s, etc, and so on for $\mathcal{N}^{(c)}$'s. $K_4(h_0)$ is a constant term which collects all $h_0$ contributions. We also have
\begin{align}\label{J4}
J_{4}(h) &=\frac{\sin^{n}(\theta)g(h)}{\dot{h}h^{n+1}\sqrt{1+h^{2}+\dot{h}^{2}}} -\sum_{i=0}\Big(N_{(n-2i-1)}+\epsilon_d\tilde{N}_{(n-2i-1)}\log h+\cdots\Big)\frac{g(h)}{h^{n-2i-1}}
\end{align}

Taking derivatives of $I_i$'s with respect to $\delta$,
\begin{align}\label{dI1toddelta}
\frac{dI_{1}}{d\delta}&=
-\frac{1}{\delta^{m+1}} \int_{h_{0}}^{0}dh J_{1}(h)+\frac{1}{\delta^{m+1}}K_1(h_0)
\nonumber\\
&-\bar{\epsilon}_{n}\frac{1}{\delta^{m+1}}\left(P_1\log(\frac{\delta}{H})+\frac{1}{2}\epsilon_d\tilde{P}_1\log^2(\frac{\delta}{H})+\cdots\right) \nonumber\\
&+\sum'_{i=0}\frac{1}{H^{m+1}}\Big(\frac{\delta}{H}\Big)^{2i-d+1}\Big(\mathcal{P}_{(n-2i+2)}+\epsilon_d\tilde{\mathcal{P}}_{(n-2i+2)}\log(\frac{\delta}{H})+\cdots\Big)
\nonumber\\
&-\frac{(\mu H)^{2\alpha}}{H^{m+1}}\Bigg\{\bar{\epsilon}_{n}\Big[\Big(\frac{\delta}{H}\Big)^{2\alpha-m}\xi_{\alpha j} b_{2j}\Big(P_1\log(\frac{\delta}{H})+\tilde{P}_1\log^2(\frac{\delta}{H})+\hat{P}_1\log^2(\frac{\delta}{H})\nonumber\\
&+\sum_{i=0}\Big(
\Big(\frac{\delta}{H}\Big)^{2\alpha+2i-m-1}(P_1b_{2i}+\epsilon_d(P_1\tilde{b}_{2i}+\tilde{P}_1b_{2i})\log(\frac{\delta}{H})+\epsilon_d(\tilde{P}_1\tilde{b}_{2i}+\hat{P}_1b_{2i})\log^2(\frac{\delta}{H}))\nonumber\\
&+\Big(\frac{\delta}{H}\Big)^{n+2i}(P_1c_{2i}+\epsilon_d(P_1\tilde{c}_{2i}+\tilde{P}_1c_{2i})\log(\frac{\delta}{H})+\epsilon_d(\tilde{P}_1\tilde{c}_{2i}+\hat{P}_1c_{2i})\log^2(\frac{\delta}{H}))+\cdots\Big)\Big]
\nonumber\\
&+ \sum_{i=0}\Bigg[\xi_{\alpha j} b_{2j}\Big(\frac{\delta}{H}\Big)^{2i-2j-1}\Big(((2i-n-1)\mathcal{P}_{(2-2i+n)}+\tilde{\mathcal{P}}_{(2-2i+n)})\log(\frac{\delta}{H})
\nonumber\\
&+((2i-n-1)\tilde{\mathcal{P}}_{(2-2i+n)}+2\hat{\mathcal{P}}_{(2-2i+n)})\log^2(\frac{\delta}{H})+\cdots\Big) \nonumber\\
&+\sum_{k=0}\Big[\Big(\frac{\delta}{H}\Big)^{2\alpha+2i+2k-d+1}\Big(b_{2k}((1-\iota)\mathcal{P}_{(\iota)}+\tilde{\mathcal{P}}_{(\iota)})
\nonumber\\
&\log(\frac{\delta}{H})\Big(b_{2k} ((1-\iota) \tilde{\mathcal{P}}_{(\iota)}+2 \hat{\mathcal{P}}_{(\iota)})+\tilde{b}_{2k} ((1-\iota) \mathcal{P}_{(\iota)}+\tilde{\mathcal{P}}_{(\iota)})\Big)\nonumber\\
&+\log^2(\frac{\delta}{H})\Big(b_{2k} (1-\iota) \hat{\mathcal{P}}_{(\iota)}+\tilde{b}_{2k} ((1-\iota) \tilde{\mathcal{P}}_{(\iota)}+2\hat{\mathcal{P}}_{(\iota)})\Big)\Big)
\nonumber\\
&+\Big(\frac{\delta}{H}\Big)^{2i+2k-1}\Big(c_{2k}((1-\iota)\mathcal{P}_{(\iota)}+\tilde{\mathcal{P}}_{(\iota)})
\nonumber\\
&\log(\frac{\delta}{H})\Big(c_{2k} ((1-\iota) \tilde{\mathcal{P}}_{(\iota)}+2 \hat{\mathcal{P}}_{(\iota)})+\tilde{c}_{2k} ((1-\iota) \mathcal{P}_{(\iota)}+\tilde{\mathcal{P}}_{(\iota)})\Big)\nonumber\\
&+\log^2(\frac{\delta}{H})\Big(c_{2k} (1-\iota) \hat{\mathcal{P}}_{(\iota)}+\tilde{c}_{2k} ((1-\iota) \tilde{\mathcal{P}}_{(\iota)}+2\hat{\mathcal{P}}_{(\iota)})\Big)\Big)+\cdots\Big]\Bigg]\Bigg\}
\end{align}
in which $\iota=n-2i+2$ and ellipses show higher log terms. We also used $m=d-n-3$ and
\begin{align}
K_{1}(h_{0})&=\bar{\epsilon}_{n}P_1\log(h_0)-\sum'_{i=0}\frac{P_{(n-2i+2)}}{n-2i+1}\frac{1}{h_{0}^{n-2i+1}}+\cdots.
\end{align}
Similarly 
\begin{align}
\frac{dI_{2}}{d\delta}&=\frac{(\mu H)^{2\alpha}}{2H^{m+1}}\Big(\frac{\delta}{H}\Big)^{2\alpha-m-1}\Big(-\int_{h_{0}}^{0}dh     J_{2}(h)+K_{2}(h_{0})\Big)
\nonumber\\
&+\frac{(\mu H)^{2\alpha}}{2H^{m+1}} \Big[-\bar{\epsilon}_{n}\Big(\frac{\delta}{H}\Big)^{2\alpha-m-1}\Big(Q_1\log(\frac{\delta}{H})+\frac{1}{2}\epsilon_d\tilde{Q}_1\log^2(\frac{\delta}{H})+\cdots\Big)
\nonumber\\
&+\sum'_{i=0}\Big(\frac{\delta}{H}\Big)^{2\alpha+2i-d+1}\Big(\mathcal{Q}_{(n-2i+2)}+\epsilon_d\tilde{\mathcal{Q}}_{(n-2i+2)}\log(\frac{\delta}{H})+\cdots\Big)\Big],
\end{align}
where 
\begin{align}
K_{2}(h_{0})&=\bar{\epsilon}_{n}Q_{1}\log(h_{0})-\sum'_{i=0}\frac{Q_{(n-2i+2)}}{n-2i+1}\frac{1}{h_{0}^{n-2i+1}}.
\end{align}
On the other hand,
\begin{align}
\frac{d{I_{3}}}{d\delta}=&-\frac{(\mu H)^{2\alpha}}{H^{m+1}}\sum_{i=0}\Big\{
\Big(\xi_{\alpha j} b_{2j}\Big(\frac{\delta}{H}\Big)^{2i-2j}\log(\frac{\delta}{H})\Big(M_{(n-2i)}+\epsilon_d\tilde{M}_{(n-2i)}\log(\frac{\delta}{H})\Big)
\nonumber\\
&+\sum_{k=0}\Big[\Big((b_{2k}+\epsilon_d\tilde{b}_{2k}\log(\frac{\delta}{H}))\Big(\frac{\delta}{H}\Big)^{2\alpha-d+2i+2k+1}
\nonumber\\
&+(c_{2k}+\epsilon_d\tilde{c}_{2k}\log(\frac{\delta}{H}))\Big(\frac{\delta}{H}\Big)^{2i+2k-1}\Big)\Big(M_{(n-2i)}+\epsilon_d\tilde{M}_{(n-2i)}\log(\frac{\delta}{H})\Big)\Big]\Big\}
\\
\frac{dI_{4}}{d\delta}=& \frac{(\mu H)^{2\alpha}}{H^{m+1}}(m-2\alpha)\Big\{\Big(\frac{\delta}{H}\Big)^{2\alpha-m-1}\Big(\int_{h_{0}}^{0}dh J_{4}(h)+K_{4}(h_0)\Big)
\nonumber\\
&+\epsilon_{n}\xi_{\alpha j} b_{2j}\Big(\frac{\delta}{H}\Big)^{2\alpha-m-1}\Big(\frac{1}{2}N_1\log^2 (\frac{\delta}{H})+\frac{1}{3}\epsilon_d\tilde{N}_1\log^3 (\frac{\delta}{H})+\cdots\Big)
\nonumber\\
&+\bar{\epsilon}_{n}\sum_{k=0}\Big(\frac{\delta}{H}\Big)^{2\alpha-m-1}\Big[b_{2k}N_{(2k)}\log (\frac{\delta}{H})+\frac{1}{2}\epsilon_d(b_{2k}\tilde{N}_{(2k)}+\tilde{b}_{2k}N_{(2k)})\log^2 (\frac{\delta}{H})
\nonumber\\
&+\frac{1}{3}\epsilon_d(\tilde{b}_{2k}\tilde{N}_{(2k)}+b_{2k}\hat{N}_{(2k)})\log^3 (\frac{\delta}{H})+\cdots\Big]
\nonumber\\
&+\zeta_{\alpha i}\sum_{k=0}\Big(\frac{\delta}{H}\Big)^{2\alpha-m-1}\Big[c_{2k}N_{(2k+d-2\alpha-2)}\log (\frac{\delta}{H})
\nonumber\\
&+\frac{1}{2}\epsilon_d(c_{2k}\tilde{N}_{(2k+d-2\alpha-2)}+\tilde{c}_{2k}N_{(2k+d-2\alpha-2)})\log^2 (\frac{\delta}{H})
\nonumber\\
&+\frac{1}{3}\epsilon_d(\tilde{c}_{2k}\tilde{N}_{(2k+d-2\alpha-2)}+c_{2k}\hat{N}_{(2k+d-2\alpha-2)})\log^3 (\frac{\delta}{H})+\cdots\Big]
\nonumber\\
&+\xi_{\alpha j} b_{2j}\sum'_{i=0}\Big(\frac{\delta}{H}\Big)^{2i-2j+2}\Big[\mathcal{N}_{(n-2i-1)}+\tilde{\mathcal{N}}_{(n-2i-1)}\log (\frac{\delta}{H})+\cdots\Big]
\nonumber\\
&+\sum'_{i,k=0}\Big[\Big(\frac{\delta}{H}\Big)^{2 \alpha+2 i+2 k-d+3} (\mathcal{N}^{(b)}_{(-2 i+n-1)}+\log (\frac{\delta}{H}) \epsilon_d\tilde{\mathcal{N}}^{(b)}_{(-2 i+n-1)}+\cdots)
\nonumber\\
&+\Big(\frac{\delta}{H}\Big)^{2 i+2 k+1} (\mathcal{N}^{(c)}_{(-2 i+n-1)}+\log (\frac{\delta}{H})\epsilon_d \tilde{\mathcal{N}}^{(c)}_{(-2 i+n-1)}+\cdots)\Big]
\end{align}
where terms with $\epsilon_n$, $\bar{\epsilon}_n$ and $\zeta_{\alpha i}$ as coefficients, indicate special cases which are extracted from summations on $i$ and we denote the excluded sum with a prime sign.

In expressions for $dI_i/d\delta$ we are interested in $1/\delta$ and $\log^\ell(\delta)/\delta$ terms which respectively correspond to log and higher log terms in the entanglement entropy. Let us consider each $I_i$'s separately. 


\begin{enumerate}

\item $I_1$:

Integrating \eqref{dI1toddelta}, possible log and higher log terms contributions are as follows,

\begin{itemize}[noitemsep,wide=0pt, leftmargin=\dimexpr\labelwidth + 2\labelsep\relax]
	\item $2\alpha=m-1=d-2\ell-2$ and $n=2\ell-2$: 
	\begin{equation}
	A_1(\ell)=-\frac{\mu^{m-1}}{2H}b_{2\ell}P_1\log^2(\frac{\delta}{H})+\cdots.
	\end{equation}
\item $2\alpha=m-2\ell$: 
	\begin{equation}
	A_2(\ell)=-\frac{(\mu H)^{2\alpha}}{H^{m}}\bar{\epsilon}_n\Big(b_{2\ell}P_1\log(\frac{\delta}{H})+\epsilon_d(\frac{1}{2}(P_1\tilde{b}_{2\ell}+\tilde{P}_1b_{2\ell})\log^2(\frac{\delta}{H})+\cdots)\Big).
	\end{equation}
\item $2\alpha=d-2\ell-2$ and $n\neq 2\ell-1$: 
	\begin{equation}
	A_3(\ell)=-\frac{(\mu H)^{2\alpha}}{2H^{m}}(n+1-2\ell)b_{2\ell}\mathcal{P}_{(n+2-2\ell)}\log^2(\frac{\delta}{H}).
	\end{equation}
\item $2\alpha=d-2\ell-2$ and $n\neq 2i-1$: 
	\begin{align}
	A_4(\ell)&=-\frac{(\mu H)^{2\alpha}}{H^{m}}\epsilon_d\sum_{i=0}^{\ell} { }'  \Big(\frac{1}{2}\log^2(\frac{\delta}{H})\Big(b_{2k} ((1-\iota) \tilde{\mathcal{P}}_{(\iota)}+2 \hat{\mathcal{P}}_{(\iota)})+\tilde{b}_{2k} ((1-\iota) \mathcal{P}_{(\iota)}+\tilde{\mathcal{P}}_{(\iota)})\Big)
	\nonumber\\
	&+\frac{1}{3}\log^3(\frac{\delta}{H})\Big(b_{2k} (1-\iota) \hat{\mathcal{P}}_{(\iota)}+\tilde{b}_{2k} ((1-\iota) \tilde{\mathcal{P}}_{(\iota)}+2\hat{\mathcal{P}}_{(\iota)})\Big)\Big).
	\end{align}
	where $\iota=n-2i+2$, $k=\ell-i$ and prime on the summation means exclusion of $i=(n+1)/2$.
\end{itemize}

\item $I_2$:

In $I_2$, the contribution is $\mu$-dependent:\\
\begin{itemize}
	\item $2\alpha=m$: 
\begin{align}
	A_5&=-\frac{\mu ^{m}}{2}\Big[\Big(-\int_{h_0}^0dhJ_2(h)+K_2(h_0)\Big)\log(\frac{\delta}{H})
\nonumber\\
&+\frac{1}{2}     \bar{\epsilon}_n\Big(Q_1\log^2(\frac{\delta}{H})+\epsilon_d(\frac{1}{3}\tilde{Q}_1\log^3(\frac{\delta}{H})+\cdots\Big)\Big]
	\end{align}
	\item $2\alpha=d-2\ell-2$ and $n\neq 2\ell-1$: 
	\begin{equation}
	A_6(\ell)=\frac{(\mu H)^{2\alpha}}{2H^m}\Big[\mathcal{Q}_{(n+2-2\ell)}\log(\frac{\delta}{H})+\epsilon_d(\frac{1}{2}\tilde{\mathcal{Q}}_{(n+2-2\ell)}\log^2(\frac{\delta}{H})+\cdots)\Big]
	\end{equation}
\end{itemize}

\item $I_3$:

In $I_3$, the contribution is $\mu$-dependent:
\begin{itemize}
	
\item $2\alpha=d-2\ell-2$: 
\begin{align}
A_7(\ell)&=-\frac{(\mu H)^{2\alpha}}{H^m}\Bigg[
\sum_{i=0}^{\ell}\Big(b_{2\ell-2i}M_{(n-2i)}\log(\frac{\delta}{H})\nonumber\\
&+\epsilon_d(\frac{1}{2}(\tilde{b}_{2\ell-2i}M_{(n-2i)}+b_{2\ell-2i}\tilde{M}_{(n-2i)})\log^2(\frac{\delta}{H})+\cdots)\Big)\Bigg]
\end{align}
\end{itemize}

\item $I_4$:
\begin{itemize}
\item $2\alpha=d-2\ell-4$: 
	 \begin{align}
	 A_8(\ell)=(m-2\alpha)\frac{(\mu H)^{2\alpha}}{H^m} \sum_{i=0}^{\ell}\Big[\mathcal{N}^{(b)}_{(n-2i-1)}
\log(\frac{\delta}{H})+\epsilon_d(\frac{1}{2}\tilde{\mathcal{N}}^{(b)}_{(n-2i-1)}
\log^2(\frac{\delta}{H})+\cdots)\Big].
	 \end{align}

\end{itemize}

\end{enumerate}


Notice that in above cases, $\ell$ values include only finite numbers of integers, $\{0,1,\dots, \ell_{max}\}$ such that the constraint $ \alpha\geq 0$ is satisfied. 

Now from the above relations we can infer that the relevant perturbed contribution to the log and higher log parts of the entanglement entropy is
\begin{align}
&2\alpha=m-1, \quad n= 2\ell-2 \quad \text{and} \quad \ell \geq 1: \nonumber\\
\label{SEE-1}
&\quad\Delta S_{EE}= \frac{L^{d-1}\Omega_{n}\tilde{H}^{m}}{2G_{N}}(A_1(\ell)+A_7(\ell)+A_8(\ell-1)),
\\
&2\alpha=m: \nonumber\\
\label{SEE-2}
&\quad\Delta S_{EE}= \frac{L^{d-1}\Omega_{n}\tilde{H}^{m}}{2G_{N}}(A_2(0)+A_5),
\\
&2\alpha=m-2\ell \quad \text{and} \quad \ell \geq 1: \nonumber\\
\label{SEE-3}
&\quad\Delta S_{EE}= \frac{L^{d-1}\Omega_{n}\tilde{H}^{m}}{2G_{N}}A_2(\ell),
\\
&2\alpha=d-2\ell-2, \quad n\neq 2i-1 \quad \text{and} \quad \ell \geq 1: \nonumber\\
\label{SEE-4}
&\quad\Delta S_{EE}= \frac{L^{d-1}\Omega_{n}\tilde{H}^{m}}{2G_{N}}\Big(A_3(\ell)+A_4(\ell)+A_6(\ell)+A_7(\ell)+A_8(\ell-1)\Big),
\\
&2\alpha=d-2: \nonumber\\
\label{SEE-6}
&\quad\Delta S_{EE}= \frac{L^{d-1}\Omega_{n}\tilde{H}^{m}}{2G_{N}}\Big(A_3(0)+A_4(0)+A_6(0)+A_7(0)\Big),
\end{align}
where $\ell=0,1,\cdots,\ell_{max}$. 
So we derived the effect of relevant perturbation with scaling dimension $\Delta$ in general for geometries in the form $c_{n}\times R^{m}$. As seen, there are logarithmic and higher logarithmic terms for some integer values of $2\alpha$. Note that many coefficients inside $A_i$'s functions are zero in even dimensions. In Appendices \ref{A} and \ref{B}, we present nonvanishing terms in $d=4$ and $d=6$.

The important results are as follows. The universal logarithmic and higher logarithmic terms appear only in special values of $\alpha$. In odd dimensions, in addition to log terms, we have double log for some special integer values of $2\alpha$. In even dimensions, there are some special values of $\alpha$ for which a power series of logarithmic terms show up.

As we see, in agreement with previous results for smooth as well as singular entangling surfaces\cite{Ref34,Ref35,Ref36,Ref37,Ref38,Ref39,Ref40,Ref41}, the relevant perturbation with scaling dimension $\Delta=\frac{d+2}{2}$ or equivalently  $\alpha=\frac{d-2}{2}$ leads to appearance of logarithmic term in entanglement entropy.

In order to have some experience for preceding calculations, we consider few examples, $d=5,6$. \\

\subsubsection{$d=5$}

Now, we consider the geometry in form $c_{1}\times R^{1}$ with $d=5$. From the previous discussion, we can summarize the contribution of relevant perturbation to the $\Delta S_{EE}^{(1)} $. From \eqref{SEE-1} to \eqref{SEE-6}, we find
\begin{itemize}[noitemsep,wide=0pt, leftmargin=\dimexpr\labelwidth + 2\labelsep\relax]
\item $\alpha=1/2$\\
\begin{align}\label{alpha-51}
\Delta S_{EE}^{(1)}&= \frac{L^{4}\Omega_{1}\tilde{H}^{1}}{2G_{N}}\Big(A_2(0)+A_5\Big)
\nonumber\\
&= -\frac{\mu L^{4}\Omega_{1}\tilde{H}^{1}}{2G_{N}}
\Big[ \frac{1}{2}\Big(-\int_{h_0}^0dhJ_2(h)+K_2(h_0)\Big)\log(\frac{\delta}{H})
\nonumber\\
&+\frac{1}{2}b_{0}P_1\log(\frac{\delta}{H})+\frac{1}{2}Q_1\log^2\frac{\delta}{H}\Big]
\end{align}
In this case, we have both logarithmic and double logarithmic terms. 

\item $\alpha =3/2$
\begin{align}\label{alpha-53}
\Delta S_{EE}^{(1)}&= \frac{L^{4}\Omega_{1}\tilde{H}^{1}}{2G_{N}}\Big(A_3(0)+A_6(0)+A_7(0)\Big)
\nonumber\\
&= \frac{L^{4}\Omega_{1}\tilde{H}^{1}}{4G_{N}}\mu^3H^2\Big[(\frac{1}{2}\mathcal{Q}_3-b_0M_{1})\log(\frac{\delta}{H})-b_0\mathcal{P}_3\log^2(\frac{\delta}{H})\Big]
\end{align}
In this case, we have both logarithmic and double logarithmic contributions to the entanglement entropy.

\end{itemize}

\subsubsection{$d=6$:}

For $d=6$ dimensions, we consider two geometries $c_{1}\times R^{2}$ and  $c_{2}\times R^{1}$ corresponding respectively to $n=1, 2$. 
Let us separate $n=1$ and $n=2$ as follows.

\begin{enumerate}
\item $n=1$ and $m=2$, $c_{1}\times R^{2}$\\
Here, we find the contribution of relevant perturbation to
 $\Delta S_{EE}^{(1)} $ in various of $\alpha $ as follows:\\

\begin{itemize}[noitemsep,wide=0pt, leftmargin=\dimexpr\labelwidth - 2\labelsep\relax]
\item  $\alpha=1$:
\begin{align}\label{alpha-1}
\Delta S_{EE}^{(1)}&=
\frac{L^{5}\Omega_{1}\tilde{H}^{2}}{2G_{N}}\Big(A_2(0)+A_5\Big)
\nonumber\\
&= -\frac{\mu^2 L^{5}\Omega_{1}\tilde{H}^{2}}{2G_{N}}
\Big[ \frac{1}{2}\Big(-\int_{h_0}^0dhJ_2(h)+K_2(h_0)\Big)\log(\frac{\delta}{H})
\nonumber\\
&+b_{0}P_1\log(\frac{\delta}{H})+\frac{1}{2}Q_1\log^2\frac{\delta}{H}\Big]
\end{align}
This result shows that in this case, we have log and double log singularities.

\item $\alpha=2$:
\begin{align}\label{alpha-63}
\Delta S_{EE}^{(1)}=
&\frac{L^{5}\Omega_{1}\tilde{H}^{2}}{2G_{N}}\Big(A_3(0)+A_4(0)+A_6(0)+A_7(0)\Big)
\nonumber\\
=& \frac{\mu^4 L^{5}\Omega_{1}\tilde{H}^{2}}{2G_{N}}\Big[(\frac{1}{2}\mathcal{Q}_{3}-b_0M_1)\log(\frac{\delta}{H})
-b_0\mathcal{P}_3\log^2(\frac{\delta}{H})\Big]
\end{align}
Again, there are both logarithmic and double logarithmic terms which contribute from the relevant perturbation to the entanglement entropy. 

\end{itemize}

\item $n=2$ and $m=1$, $c_{2}\times R^{1}$\\
In this geometry, we find for various values of $\alpha$:

\begin{itemize}[noitemsep,wide=0pt, leftmargin=\dimexpr\labelwidth - 2\labelsep\relax]
\item  $\alpha=1/2$:
\begin{align}\label{alpha-1}
\Delta S_{EE}^{(1)}=
&\frac{L^{5}\Omega_{2}\tilde{H}^{1}}{2G_{N}}A_5
\nonumber\\
=&- \frac{\mu L^{5}\Omega_{2}\tilde{H}^{1}}{2G_{N}}
\Big[ \frac{1}{2}\Big(-\int_{h_0}^0dhJ_2(h)+K_2(h_0)\Big)\log(\frac{\delta}{H})
\nonumber\\
&+\frac{1}{2}Q_1\log^2\frac{\delta}{H}\Big]
 \end{align}

\item  $\alpha=1$:
\begin{align}
\Delta S_{EE}^{(1)}=
&\frac{L^{5}\Omega_{2}\tilde{H}^{1}}{2G_{N}}\Big(A_3(1)+A_4(1)+A_6(1)+A_7(1)+A_8(0)\Big)
\nonumber\\
=& \frac{\mu^2H L^{5}\Omega_{2}\tilde{H}^{1}}{4G_{N}}\Big[(\mathcal{Q}_2-2b_2M_2-2b_0M_0-\mathcal{N}^{(b)}_1)\log(\frac{\delta}{H})
\nonumber\\
&+(3\tilde{b}_2\mathcal{P}_4-b_2\mathcal{P}_2-\tilde{b}_2M_2-\tilde{\mathcal{N}}^{(b)}_{1})\log^2(\frac{\delta}{H})+\cdots\Big]
\end{align}

\item  $\alpha=2$:
\begin{align}
\Delta S_{EE}^{(1)}=
&\frac{L^{5}\Omega_{2}\tilde{H}^{1}}{2G_{N}}\Big(A_3(0)+A_4(0)+A_6(0)+A_7(0)\Big)
\nonumber\\
=& \frac{\mu^4H^3 L^{5}\Omega_{2}\tilde{H}^{1}}{2G_{N}}\Big\{\Big[ 
\frac{1}{2}\mathcal{Q}_4-b_0M_2\Big]\log(\frac{\delta}{H})
\nonumber\\
&+\frac{1}{2}\Big[3(\tilde{b}_0-b_0)\mathcal{P}_{4}-\tilde{b}_0M_2
\Big]\log^2(\frac{\delta}{H})+\cdots\Big\}
 \end{align}

\end{itemize}

\end{enumerate}
As seen in $d=6$, we have few discrete values of $\alpha$ for which we have universal  log terms.

\subsection{ Cone $c_{n}$ }
\label{sec3.1}
In this subsection, we study the holographic entanglement entropy of the cone $c_{n}$ in a relevant perturbation of a conformal field theory using an asymptotic anti-de Sitter space time background. Indeed, this case is a special case of $c_n\times R^m$ with $m=0$. A new logarithmic term comes from integrating the first two lines of \eqref{dI1toddelta} with $m=0$.

Now looking for $1/\delta$ and $\log^k(\delta)/\delta$ terms in expressions for $dI_i/d\delta$. Let us consider each $I_i$'s separately. 
\begin{enumerate}[noitemsep,wide=0pt, leftmargin=\dimexpr\labelwidth + 2\labelsep\relax]
\item $I_1$:
\begin{itemize}[noitemsep,wide=0pt, leftmargin=\dimexpr\labelwidth + 2\labelsep\relax]

\item independent of $\mu$ and $\alpha$:
\begin{equation}
B_1=\Big(K_1(h_0)-\int_{h_0}^0dh J_1(h)\Big)\log(\frac{\delta}{H})
-\frac{1}{2}\epsilon_dP_1\log^2(\frac{\delta}{H})
\end{equation}
in which  we replaced $\epsilon_d=\bar{\epsilon}_n$, since $d=n+3$. So a double log term appears in even $d$.
\item $2\alpha=d-2\ell-2$ and $n\neq 2\ell-1$: 
	\begin{equation}
	B_3(\ell)=-\frac{(\mu H)^{2\alpha}}{2}(n+1-2\ell)b_{2\ell}\mathcal{P}_{(n+2-2\ell)}\log^2(\frac{\delta}{H}).
	\end{equation}
\item $2\alpha=d-2\ell-2$ and $n\neq 2i-1$: 
	\begin{align}
	B_4(\ell)&=-\frac{(\mu H)^{2\alpha}}{2}\sum_{i=0}^{\ell}{ }' \Big(\frac{1}{2}\log^2(\frac{\delta}{H})\Big(b_{2k} ((1-\iota) \tilde{\mathcal{P}}_{(\iota)}+2 \hat{\mathcal{P}}_{(\iota)})+\tilde{b}_{2k} ((1-\iota) \mathcal{P}_{(\iota)}+\tilde{\mathcal{P}}_{(\iota)})\Big)
	\nonumber\\
	&+\frac{1}{3}\log^3(\frac{\delta}{H})\Big(b_{2k} (1-\iota) \hat{\mathcal{P}}_{(\iota)}+\tilde{b}_{2k} ((1-\iota) \tilde{\mathcal{P}}_{(\iota)}+2\hat{\mathcal{P}}_{(\iota)})\Big)\Big).
	\end{align}
\end{itemize}

\item $I_2$:

\begin{itemize}[noitemsep,wide=0pt, leftmargin=\dimexpr\labelwidth + 2\labelsep\relax]

	\item $2\alpha=d-2\ell-2$ and $n\neq 2\ell-1$: 
	\begin{equation}
	B_6(\ell)=\frac{(\mu H)^{2\alpha}}{2}\Big[\mathcal{Q}_{(n+2-2\ell)}\log(\frac{\delta}{H})+\epsilon_d(\frac{1}{2}\tilde{\mathcal{Q}}_{(n+2-2\ell)}\log^2(\frac{\delta}{H})+\cdots)\Big]
	\end{equation}
\end{itemize}

\item $I_3$:

\begin{itemize}[noitemsep,wide=0pt, leftmargin=\dimexpr\labelwidth + 2\labelsep\relax]
	
\item $2\alpha=d-2\ell-2$: 
\begin{align}
B_7(\ell)&=-(\mu H)^{2\alpha}\sum_{i=0}^{\ell}\Big(b_{2\ell-2i}M_{(n-2i)}\log(\frac{\delta}{H})\nonumber\\
&+\epsilon_d(\frac{1}{2}(\tilde{b}_{2\ell-2i}M_{(n-2i)}+b_{2\ell-2i}\tilde{M}_{(n-2i)})\log^2(\frac{\delta}{H})+\cdots)\Big)
\end{align}
\end{itemize}

\item $I_4$:

\begin{itemize}[noitemsep,wide=0pt, leftmargin=\dimexpr\labelwidth + 2\labelsep\relax]
\item $2\alpha=d-2\ell-4$: 
	 \begin{align}
	 B_8(\ell)=-2\alpha(\mu H)^{2\alpha} \sum_{i=0}^{\ell}\Big[\mathcal{N}^{(b)}_{(n-2i-1)}
\log(\frac{\delta}{H})+\epsilon_d(\frac{1}{2}\tilde{\mathcal{N}}^{(b)}_{(n-2i-1)}
\log^2(\frac{\delta}{H})+\cdots)\Big].
	 \end{align}

\end{itemize}

\end{enumerate}


Notice that in above cases, $\ell$ is an integer in the range $0\leq \ell \leq \ell_{max}$ such that the constraint $ \alpha\geq 0$ is satisfied. In contrast to the $c_n\times R^m$ case, we removed $2\alpha=m=0$ cases as noninteresting ones and added $B_1$ as a new ingredient. However, the later does not contribute to the relevant perturbation of the entanglement entropy.


Now from the above relations we can infer that the relevant perturbed contribution to the log and higher log parts of the entanglement entropy is
\begin{align}
&2\alpha=d-2\ell-2\quad \text{and} \quad n\neq 2\ell-1 \quad \text{and} \quad \ell \geq 1: \nonumber\\
\label{SEE2-3}
&\quad\Delta S_{EE}= \frac{L^{d-1}\Omega_{n}}{2G_{N}}\Big(B_3(\ell)+B_4(\ell)+B_6(\ell)+B_7(\ell)+B_8(\ell-1)\Big),
\\
&2\alpha=d-2: \nonumber\\
\label{SEE2-5}
&\quad\Delta S_{EE}= \frac{L^{d-1}\Omega_{n}}{2G_{N}}\Big(B_3(0)+B_4(0)+B_6(0)+B_7(0)\Big),
\end{align}
where $\ell=0,1,\cdots,\ell_{max}$. 
So we derived the effect of relevant perturbation with scaling dimension $\Delta$ in general for geometries in the form $c_{n}$. As seen, there are logarithmic and double logarithmic terms for some integer values of $2\alpha$. 

In $d=4$ dimension, only $2\alpha=d-2=2$ contributes to a universal logarithmic term as,
\begin{align}
\Delta S_{EE}&= \frac{L^{d-1}\Omega_{n}}{2G_{N}}\Big(B_3(0)+B_4(0)+B_6(0)+B_7(0)\Big) \nonumber\\
&= \frac{L^{3}\Omega_{1}}{2G_{N}}(\mu H)^2\Big[\Big(\frac{1}{2}\mathcal{Q}_3-b_0M_1-2\mathcal{N}^{(b)}_0\Big)\log(\frac{\delta}{H})
\nonumber\\
&+\Big((\frac{1}{2}\tilde{b}_0-b_0)\mathcal{P}_3+\frac{1}{2}\tilde{b}_0M_1\Big)\log^2(\frac{\delta}{H})\Big]
\end{align}
where explicit values of coefficients are given in appendices.


\subsection{Crease $k\times R^{m}$}
\label{sec1}
In this subsection, we will study the holographic entanglement entropy of crease region $k\times R^{m}$ in asymptotic anti-de Sitter space time background. This is similar to the $c_n\times R^m$ in section (\ref{sec3.3}) with $n=0$. 
As before, we choose the bulk metric in the form $(1.6)$. Similar to pure AdS backgrounds, we pick $\sigma=(\rho,\theta,x^{i},z)$ as the induced coordinates over the minimal surface, and parameterize the bulk minimal surface as $ \rho=\rho(z,\theta)$. So the induced metric on the bulk minimal surface in time slice $t=0$ becomes
\begin{equation}
 ds ^{2}= \frac{L^{2}}{z^{2}}\Big( \big(\rho^{'2}+\frac{1}{f(z)}\big)dz^{2}+2\rho'\dot{\rho}dzd\theta+(\dot{\rho}^{2}+\rho^{2})
 d\theta^{2}+\sum_{i=1}^{m}(dx^{i})^{2}\Big),
\end{equation}
where $\dot{\rho}= \partial_{\theta}\rho$, $\rho'=\partial_{z}\rho$. 
Now, the entanglement entropy is given by
 \begin{equation}
   S_{EE}=\frac{1}{4G_{N}}\int d\sigma\sqrt{\gamma}=\frac{L^{d-1}\tilde{H}^{d-3}}{4G_{N}}\int_{\delta}^{z_{m}}dz\int_{\frac{-\Omega}{2}+\epsilon}^{\frac{\Omega}{2}-\epsilon}
   d\theta\frac{1}{z^{d-1}\sqrt{f}}\sqrt{(f\rho^{'2}+1)\rho^{2}+\dot{\rho}^{2}}
 \end{equation}
 from which the equation of motion for $\rho$ can be derived and is equivalent to
  \eqref{eq-rho} with replacing $n=0$.
 
 For small deformation, we can make the following ansatz
\begin{equation}
  \rho(z,\theta)=\rho_{0}+\delta\rho=\frac{z}{h(\theta)}+\delta f g(\theta)z,\qquad  f=1+\delta f
\end{equation}
with $\delta f(z)=\mu^{2\alpha} z^{2\alpha}$. In the first order of perturbation, using  equation of motion for $\rho_{0}$, we will find the following entropy functional
\begin{align}\label{SEE-K}
&S_{EE}=\frac{L^{d-1}\tilde{H}^{d-3}}{4G_{N}}\int_{\delta}^{z_{m}}dz \int_{\frac{-\Omega}{2}+\epsilon}^{\frac{\Omega}{2}-\epsilon}d\theta
\Big[\frac{R_0}{z^{d-1}}
-\frac{(\rho_{0}^{2}+\dot{\rho}^{2}) \delta f}{2z^{d-1}R_0}
+\partial_{\theta}\Big(
\frac{\dot{\rho_{0}}\delta\rho}{z^{d-1}R_0}\Big)
+\partial_{z}\Big(
\frac{\rho_{0}^{2}\rho_{0}'\delta\rho}
{z^{d-1}R_0}\Big)\Big]
\end{align}
where $R_0=\sqrt{\rho_0^2(1+\rho_0'^2)+\dot{\rho}_0^2}$.
Equation \eqref{SEE-K} is equivalent to \eqref{eq5} with $n=0$. Now by substituting the ansatz $\rho(z,\theta)=\rho_{0}+\delta\rho=\frac{z}{h(\theta)}+z\delta fg(\theta)$ in the above functional we reach to following functional:
\begin{equation}
S_{EE}=\frac{L^{d-1}\tilde{H}^{d-3}}{2G_{N}}\Big (I_{1}+I_{2}+I_{3}+I_{4}\Big),
\end{equation}
where $I_{1}$, $I_{2}$, $I_{3}$, and $I_{4}$ defined as:
\begin{align}\label{K-integrals}
I_{1} &=\int_{\delta}^{z_{m}}\frac{dz}{z^{d-2}} \int_{0}^{\frac{\Omega}{2}-\epsilon}d\theta
\frac{\sqrt{1+h^{2}+\dot{h}^{2}}}{h^{2}}
\nonumber\\
 &=\int_{\delta}^{z_{m}}\frac{dz}{z^{d-2}} \int_{h_{0}}^{h_{c}}dh
\frac{\sqrt{1+h^{2}+\dot{h}^{2}}}{\dot{h}h^{2}}
\nonumber\\
I_{2}&=\int_{\delta}^{z_{m}}dz\frac{\delta f}{2z^{d-2}} \int_{0}^{\frac{\Omega}{2}-\epsilon}d\theta
\frac{-(h^{2}+\dot{h}^{2}) }{h^{2}\sqrt{1+h^{2}+\dot{h}^{2}}}
\nonumber\\
&=\int_{\delta}^{z_{m}}dz\frac{\delta f}{2z^{d-2}} \int_{h_{0}}^{h_{c}}dh
\frac{-(h^{2}+\dot{h}^{2}) }{\dot{h}h^{2}\sqrt{1+h^{2}+\dot{h}^{2}}},
\nonumber\\
I_{3}&=\int_{\delta}^{z_{m}}dz \int_{0}^{\frac{\Omega}{2}-\epsilon}d\theta
\partial_{\theta}\Big(
\frac{\dot{\rho_{0}}\delta\rho}{z^{d-1}\sqrt{\rho_{0}^{2}(1+\rho_{0}'^{2})+\dot{\rho_{0}}^{2}}}\Big)
\nonumber\\
&=\int_{\delta}^{z_{m}}dz\frac{\delta f}{z^{d-2}}\frac{-\dot{h}g(\theta)}{\sqrt{1+h^{2}+\dot{h}^{2}}}|_{h=h_{c}}
\nonumber\\
I_{4}&=\int_{\delta}^{z_{m}}dz \int_{0}^{\frac{\Omega}{2}-\epsilon}d\theta
\partial_{z}\Big(
\frac{\rho_{0}^{2}\rho_{0}'\delta\rho}
{z^{d-1}\sqrt{\rho_{0}^{2}(1+\rho_{0}'^{2})+\dot{\rho_{0}}^{2}}}\Big)
\nonumber\\
&=\int_{\delta}^{z_{m}}dz\frac{z\delta f'-(d-3)\delta f}{z^{d-2}} \int_{h_{0}}^{h_{c}}dh
\frac{g}{\dot{h}h\sqrt{1+h^{2}+\dot{h}^{2}}}
\end{align}
where we have changed the integration variable to $h(\theta)$. We have also defined
$h_{0}=h(0)$ and $h_{c}=h(\frac{\Omega}{2}-\epsilon)$ and used $\dot{h}_{0}(0)=0$ in the getting the boundary terms.

By variation of $S_{EE}$, we derived the equation of motions for $h$ and $g$ which are the same as \eqref{eq-h} and \eqref{eq4} with $n=0$.
Then to extract the logarithmic divergence, we must consider the asymptotic behaviors of $h$  and $g$, where the later can be rewritten from \eqref{kRm-g},
\begin{align}\label{kRm-g}
g(h)&=\sum_{i=0}e_{2i}h^{2i-2\alpha-1}+\sum_{i=0}h^{2i-1}\Big[d_{2i}+\epsilon_d(\tilde{d}_{2i}\log h+\cdots)\Big].
\end{align}
where $e$'s and $d$'s are similar to $c$'s and $b$'s in section \ref{sec3.3} and are derived for few examples in Appendix \ref{A}. Then it follows that
\begin{align}
h_{c}(\delta)&=\frac{\delta}{H}+(\mu H)^{2\alpha}\Big(\frac{\delta}{H}\Big)^{2\alpha+2}g(\frac{\delta}{H})
\end{align}

Now we analyze the divergence of each one of the integrands \eqref{K-integrals}. Near the boundary, they in the asymptotic limit behaves as
\begin{align}
\frac{\sqrt{1+h^{2}+\dot{h}^{2}}}{\dot{h}h^{2}} &\sim -\frac{1}{h^{2}}-\frac{1}{2}K_{d}^{2}h^{2(d-2)}+\cdots ,
\\
-\frac{h^{2}+\dot{h}^{2}}{\dot{h}h^{2}\sqrt{1+h^{2}+\dot{h}^{2}}}&\sim \frac{1}{h^{2}}-\frac{1}{2}K_{d}^{2}h^{2(d-2)}+\cdots ,
\\
\frac{-\dot{h}g}{\sqrt{1+h^{2}+\dot{h}^{2}}}&\sim g(1-\frac{1}{2}K_{d}^{2}h^{2(d-1)}+\cdots),
\\
\frac{g}{\dot{h}h\sqrt{1+h^{2}+\dot{h}^{2}}}&\sim g(-K_{d}^{2}h^{2d-3}+(d-1)K_{d}^{2}h^{2d-1}+\cdots).
\end{align}
where $K_{d}$ is the conserved quantity given in \eqref{Kd}. Hence we can isolate the divergent part of integrals in the following way:
\begin{align}
I_{1}&=\int_{\delta}^{z_{m}}\frac{dz}{z^{d-2}} \int_{h_{0}}^{h_{c}}dh
\frac{\sqrt{1+h^{2}+\dot{h}^{2}}}{\dot{h}h^{2}}
\nonumber\\
&=\int_{\delta}^{z_{m}}\frac{dz}{z^{d-2}} \int_{h_{0}}^{h_{c}}dh
[\frac{\sqrt{1+h^{2}+\dot{h}^{2}}}{\dot{h}h^{2}}+\frac{1}{h^{2}}]+\int_{\delta}^{z_{m}}\frac{dz}{z^{d-2}}
(\frac{1}{h_{c}}-\frac{1}{h_{0}})
\end{align}
Now we differentiate it with respect to $UV$ cut-off $\delta$ and look for various divergent terms. We find
\begin{align}
\frac{dI_{1}}{d\delta}&=-\frac{1}{\delta^{d-2}}\int_{h_{0}}^{0}dh
\Big[\frac{\sqrt{1+h^{2}+\dot{h}^{2}}}{\dot{h}h^{2}}+\frac{1}{h^{2}}\Big]
-\frac{H}{\delta^{d-1}}+\frac{1}{h_{0}}\frac{1}{\delta^{d-2}}\nonumber\\
&+\frac{(\mu H)^{2\alpha}}{H^{d-2}}\Big\{
\sum_{i=0}\Big[\Big(\frac{\delta}{H}\Big)^{2i-d+1}e_{2i}+\Big(\frac{\delta}{H}\Big)^{2\alpha+2i-d+1}(d_{2i}+\tilde{d}_{2i}\log(\frac{\delta}{H})+\cdots)\Big]\Big\}
\end{align}
Similarly, we can find that
\begin{align}
{I_{2}}&=\int_{\delta}^{z_{m}}dz\frac{\delta f}{2z^{d-2}} \int_{h_{0}}^{h_{c}}dh
\frac{-(h^{2}+\dot{h}^{2}) }{\dot{h}h^{2}\sqrt{1+h^{2}+\dot{h}^{2}}}
\nonumber\\
&=-\int_{\delta}^{z_{m}}dz\frac{\delta f}{2z^{d-2}}\int_{h_{0}}^{h_{c}}dh\Big[\frac{(h^{2}+\dot{h}^{2}) }{\dot{h}h^{2}\sqrt{1+h^{2}+\dot{h}^{2}}}+\frac{1}{h^{2}}\Big]
-\int_{\delta}^{z_{m}}dz\frac{\delta f}{2z^{d-2}}(\frac{1}{h_{c}}-\frac{1}{h_{0}})
\end{align}
Then,
\begin{align}
\frac{dI_{2}}{d\delta}&=\frac{\mu^{2\alpha}}{2\delta^{d-2\alpha-2}}\Big[\int_{h_{0}}^{0}dh\Big(\frac{(h^{2}+\dot{h}^{2}) }{\dot{h}h^{2}\sqrt{1+h^{2}+\dot{h}^{2}}}+\frac{1}{h^{2}}\Big)-\frac{1}{h_0}\Big]+\frac{\mu^{2\alpha}}{2}\frac{H}{\delta^{d-2\alpha-1}}
+\cdots.
\end{align}
For boundary term we find that
\begin{align}
\frac{d{I_{3}}}{d\delta}&=-\frac{(\mu H)^{2\alpha}}{H^{d-2}}\Big\{\sum_{i=0}\Big(\frac{\delta}{H}\Big)^{2i-d+1}\Big[e_{2i}\Big]
\nonumber\\
&+\sum_{i=0}\Big(\frac{\delta}{H}\Big)^{2\alpha+2i-d+1}\Big[d_{2i}+\epsilon_d(\tilde{d}_{2i}\log  (\frac{\delta}{H})+\cdots)\Big]\Big\}.
\end{align}
and for $I_4$, 
\begin{align}
\frac{d{I_{4}}}{d\delta}&= \frac{(\mu H)^{2\alpha}}{H^{d-2}}(2 \alpha-d+3)K_d^2\sum_{i=0}\Big[
\frac{1}{2\alpha-2d-2i+3} \Big(\frac{\delta}{H}\Big)^{d+2 i-1}e_{2i}\nonumber\\
&+\frac{ \Big(\frac{\delta}{H}\Big)^{2 \alpha+d+2 i-1} (d_{2i}(2 \alpha + d + 2 i) \beta+\tilde{d}_{2i} (3 - 2 \alpha - 3 d - 4 i +(2 \alpha + d + 2 i) \beta \log (\frac{\delta}{H} )))}{\beta^2 (2 \alpha+d+2 i)^2}
\Big]
\end{align}
where $\beta= (2d+2i-3)$. 
We can isolate the logarithmic terms for each of $I_1$ to $I_4$ as follows:
\begin{enumerate}[noitemsep,wide=0pt, leftmargin=\dimexpr\labelwidth + 2\labelsep\relax]
\item $I_1$:
\begin{itemize}[noitemsep,wide=0pt, leftmargin=\dimexpr\labelwidth + 2\labelsep\relax]

\item $d=2\ell+2$:
\begin{equation}
D_1(\ell)=\frac{(\mu H)^{2\alpha}}{H^{d-2}}e_{2\ell}\log(\frac{\delta}{H})
\end{equation}
	\item $2\alpha=d-2\ell-2$: 
	\begin{equation}
	D_2(\ell)=\frac{(\mu H)^{2\alpha}}{H^{d-2}}\Big(d_{2\ell}\log(\frac{\delta}{H})+\epsilon_d(\frac{1}{2}\tilde{d}_{2\ell}\log^2(\frac{\delta}{H})+\cdots)\Big).
	\end{equation}

\end{itemize}

\item $I_2$:

\begin{itemize}[noitemsep,wide=0pt, leftmargin=\dimexpr\labelwidth + 2\labelsep\relax]

	\item $2\alpha=d-2$: 
	\begin{equation}
	D_3=\frac{\mu^{d-2}}{2}\log(\frac{\delta}{H})
	\end{equation}
\end{itemize}

\item $I_3$:

\begin{itemize}[noitemsep,wide=0pt, leftmargin=\dimexpr\labelwidth + 2\labelsep\relax]
	
\item $2\alpha=d-2\ell-2$: 
\begin{align}
D_4(\ell)&=-\frac{(\mu H)^{2\alpha}}{H^{d-2}}\Big(d_{2\ell}\log(\frac{\delta}{H})+\epsilon_d(\frac{1}{2}\tilde{d}_{2\ell}\log^2(\frac{\delta}{H})+\cdots)\Big)
\end{align}

\item $d=2\ell+2$: 
\begin{align}
D_5(\ell)&=-\frac{(\mu H)^{2\alpha}}{H^{d-2}}e_{2\ell}\log(\frac{\delta}{H})
\end{align}

\end{itemize}

\end{enumerate}
There is no log contribution from $I_4$ to the entanglement entropy. It is interesting that $D_1$ and $D_2$ are respectively, the opposite of $D_5$ and $D_4$. So after collecting all terms, the only universal logarithmic contribution to the entanglement entropy is acheived when $2\alpha=d-2$ as follows
\begin{align}\label{SEE-333}
\Delta S_{EE}^{(1)}&= \frac{L^{d-1}\tilde{H}^{d-3}}{2G_{N}}\Big(D_3+D_4(0)\Big), 
\end{align}
Note that $2\alpha=d-2$ is equivalent to $\Delta=(d+2)/2$.

\section{Conclusion}
\noindent

In this paper, we considered the effect of the relevant perturbation of a CFT on the entanglement entropy and identified the logarithmic as well as double logarithmic terms that may appear in the holographic entanglement entropy of a various higher dimensional singular surfaces.

 In the context of quantum filed theory, the entanglement entropy is $UV$ divergent due to the short range correlations in the vicinity of the entangling surface. In general, the $EE$ contains power law divergent terms. The coefficients of these terms are scheme dependent and sensitive to the details of the $UV$ regulator $\delta$. But, there are subheading contributions that are logarithmic, and their coefficients are universal which characterizing the underlying theory. 
 
 There are various logarithmic terms which may appear in the $EE$, and each of them has a separate source. In general, the appearance of these terms depends on the dimension of space time, the geometry of entangling surface, and features of the underlying theory. For example, in the vacuum state of the even dimensional $CFT$ with a smooth entangling surface, the $EE$ contains a logarithmic term such that its coefficient is some linear combination of central charges that appearing in the trace anomaly of the $CFT$. In an odd dimensional $CFT$ there is no such logarithmic contribution.
 
 Another kind of the logarithmic term that may be appear is due to a singularity of the entangling surface and independent of dimension of space time. It was shown in \cite{Ref28,Ref30,Ref32} and section $2$ of the present paper, the singularity of the entangling surface induces a logarithmic or double logarithmic terms, depending on the dimension of space time. As an example, for $3d$ $CFT$s, there appears a universal logarithmic term of the entanglement entropy which its coefficient corresponds to the central charge of the theory and is universal for general three dimensional $CFT$s.

 As was noted out in the introduction, similar universal logarithmic contribution appearers due to the relevant perturbation of a $CFT$. The holographic and field theoretic calculations shows that the $EE$ receives a logarithmic correction when the underlying $CFT$ is perturbed with the scaling dimension $\Delta=(d+2)/2$. As was shown in \cite {Ref35} for smooth case and in \cite{Ref41} for singular case in $d=3$, a new logarithmic term appears which corresponds to a relevant perturbation of the $CFT$ with a coefficient depending on the scaling dimension $\Delta$ of the relevant operator.

The importance of these universal contribution lies in that the contribution of these terms is universal in the sense that is independent of the precise details of the $UV$ regulator, and their coefficients are encode universal data which help us to probe the characteristics of the underlying theory.

In order to consider the effect of the relevant perturbation in higher dimensional singularity, we chose the singular regions in the form $c_{n}$, $k\times R^{m}$ and $c_{n}\times R^{m}$. We observed that as well as a new logarithmic term, which is unique to relevant perturbation, the double logarithmic term appear that depend on the scaling dimension of the relevant operator $\Delta$.

In the case of higher dimensional singular surfaces we found that a perturbation with relevant operator with some scaling dimension $\Delta=$ yields a universal contribution in the double logarithmic form. The explicit computations in $d=5,6$ confirm of our claim.

Due to universality and scheme independent of logarithmic terms, we only identified these terms and released the power law divergent terms may be appear in the form  $1/\delta^{q}$ or  $\frac{log^{p}\delta}{\delta^{q}}$. 

All previous computations in smooth case \cite{Ref35} and our results in the singular case in $d=3$ \cite{Ref41}, and higher dimensions expresses this point that the effect of relevant perturbation is producing the logarithmic or double logarithmic terms, in descrete values of $\alpha$, with respect to the scaling dimension of the relevant operator.

Finally, it would be interesting to investigate these divergence structures in higher derivative theories similar to \cite{Ref20,Ref51,Ref52,Ref53}.

\vspace{1cm}
\noindent {\large {\bf Acknowledgment} }  MG would like to thank Sepideh Mohammadi for encouragement and valuable comments.

\vspace{1cm}
\noindent {\large {\bf Note added} } Before  submission of this article, we received article \cite{Bakhshaei:2019ope} which has some overlap with this work.

\appendix
\section{Solving Equations of $y$ and $g$}\label{A}
To solve $y$ equation, firstly note that it is symmetric under $h\rightarrow -h$, Thus the series solution of $y$ includes either odd or even powers of $h$. However, we know $y(0)=\sin(\Omega/2)$, so we have even expansion:
\begin{align}\label{Y}
y(h)=\sin\frac{\Omega}{2}+a_2 h^2+a_4h^4+\cdots
\end{align}
Substituting in \eqref{y-equation}, and trying to find $a_{2k}$ coefficient, one finds the lowest term in which it appears, 
\begin{align}
\Big[\Big(\frac{1}{2}k(2k-d)\csc(\frac{\Omega}{2})\sin^2(\Omega)\Big) a_{2k}+f(a_{2i})\Big] h^{k-1}+\cdots=0
\end{align} 
where $f(a_{2i})$ is a known combination of $a_{2i}$'s with $i<k$. The term in the bracket should be set zero to find $a_{2k}$ in terms of lower order terms $a_{2i}$'s. However, it is singular for $d=2k$. We therefore  use \eqref{Y} only for odd dimensions and find few coefficients as follows,
\begin{align}\label{y-series1}
&a_0=\sin\frac{\Omega}{2}
\\
&a_2=\frac{n\cos(\frac{\Omega}{2})\cot(\frac{\Omega}{2})}{4-2d}
\\ \label{y-series3}
&a_4=-\frac{n\csc^{5}(\frac{\Omega}{2})\Big[\Big((d-2)^{2}-2n\Big)n+\Big(2(d-2)^{2}-(d-2)dn+2n^{2}\Big)\sin^{2}(\frac{\Omega}{2})\Big]\sin^{2}(\Omega)}{32(d-4)(d-2)^{3}}.
\end{align}

For even dimensions, $d=2k$, we consider the following expansion:
\begin{align}\label{y-even-d}
y(h)&=\sum_{i=0}a_{2i}h^{2i}+\log h\sum_{i=k}\tilde{a}_{2i}h^{2i}+(\log h)^2\sum_{i=k}\hat{a}_{2i}h^{2i}+\cdots.
\end{align}
This is a double sum in powers of $h$ and $\log h$. 
For $d=4$, we have $n=1$ and find
\begin{align}\label{log-y}
y&=\sin(\frac{\Omega}{2})-\frac{\cos(\frac{\Omega}{2})\cot(\frac{\Omega}{2})}{4}h^{2}+\cdots
\nonumber\\
&+\log h\Big[\frac{1}{64}(3-\cos\Omega )\csc(\frac{\Omega}{2})\cot ^{2}(\frac{\Omega}{2})h^{4}+\cdots\Big]
\nonumber\\
&+\left(\log h\right)^2\Big[-\frac{(5 \cos\Omega +9) \left(\cos \left(\frac{3 \Omega }{2}\right)-5 \cos (\frac{\Omega }{2})\right)^2 \csc ^7(\frac{\Omega }{2})}{131072}h^8+\cdots\Big]
\nonumber\\
&+\left(\log h\right)^3\big[\frac{(3-\cos\Omega )^3 \cot ^4(\frac{\Omega }{2}) \csc ^5(\frac{\Omega }{2})}{81920}h^{10}+\cdots\Big]+\cdots.
\end{align}
For $d=6$, $a_0$ to $a_4$ are the same as \eqref{y-series1} to \eqref{y-series3} and the first logarithmic term is
\begin{align}\label{log-y6}
\tilde{a}_6&=\frac{n(n-4)}{98304}\Big(-576 + 176 n + 4 n^2 + 21 n^3 + 
 4 (192 - 96 n - 16 n^2 + 7 n^3) \cos\Omega 
\nonumber\\
&+ (-4 +  n)^2 (-12 + 7 n) \cos(2 \Omega)\Big)\cot^2(\frac{\Omega}{2}) \csc^3(\frac{\Omega}{2})
\end{align}

Now we can solve $g$ equation. Firstly, rewrite \eqref{g-equation} in the following standard form,
\begin{equation}\label{inhomog}
g''(h)+G_1(h)g'(h)+G_2(h)g(h)=G(h)
\end{equation}
To find the general solution, we firstly need to solve the homogeneous equation by setting $G(h)=0$. 
\begin{equation}\label{homog}
g''(h)+G_1(h)g'(h)+G_2(h)g(h)=0
\end{equation}
Since $G_1(h)$ and $G_2(h)$ depend on $y(h)$, so we need to consider odd and even dimensions separately. For odd dimensions, we take a series solution as,
\begin{align}
g(h)=h^x (c_0 + c_1 h + c_2 h^2 + \cdots)
\end{align}  
Plugging into the homogeneous equation \eqref{homog}, in the lowest order we solve for $x$ to find 
\begin{align}
x_1=-2\alpha-3 , \qquad x_2=-2\alpha-3+d.
\end{align}
Now consider $x=x_1$, a series solution can be found as 
\begin{align}
g_1(h)&=\frac{1}{h^{2\alpha+3}}\Bigg[1+\frac{h^2}{(d-4) (d-2)^2 (-1 + \cos\Omega)}\Bigg( - (1 + a) (-4 + d) (-2 + 
    d) n  
\nonumber\\
& -(2 + a (1 + 2 a) (d-4) - d) (d-2) - n^2 + \Big((2 + a (1 + 2 a) (d-4) - d) (d-2) 
\nonumber\\
&+ (d-2) (d-2 + a (d-4) ) n - n^2\Big)  \cos\Omega\Bigg) +\cdots\Bigg].
\end{align}
For $x=x_2$,  
\begin{align}
g_2(h)&=\frac{1}{h^{2\alpha+3-d}}\Bigg[1+\frac{h^2}{2(d^2-2d-8) (d-2)^2 (-1 + \cos\Omega)}\Bigg( 4 a^2 (d-4) (d-2)^2 + d^5 
\nonumber\\
&- 2 a (d-4) (d-2)^2 (2 d - n-1) - d^4 (9 + n) - 
 d^3 (n^2 - 10 n-30) - 4 ( n^2 + 8 n-4) 
 \nonumber\\
 &+ 2 d ( n^2 + 28 n+4) + 
 d^2 ( 3 n^2 - 36 n -40)-\Big(4 a^2 (-4 + d) (-2 + d)^2 + d^5
\nonumber\\
& - 
   2 a (d-4) (d-2)^2 ( 2 d - n-1 ) + 4 (n-2)^2 - 
   d^4 (9 + n) + d (8 + 24 n - 2 n^2) 
\nonumber\\
&+ d^3 (n^2 + 6 n + 30) - 
   d^2 (3 n^2 + 16 n +40 )\Big)   \cos\Omega\Bigg) +\cdots\Bigg]
\end{align}
For a generic $\alpha$, these two solutions are linearly independent. So we can find the general solution to the inhomogeneous equation \eqref{inhomog} as 
\begin{align}\label{u1g1-u2g2}
g(h)=u_1(h)g_1(h)+u_2(h)g_2(h)
\end{align}
where
\begin{align} \label{u1-u2}
u_1&=-\int \frac{g_2(h) G(h)}{W[g_1,g_2]}dh+C_1
\nonumber\\
u_2&=\int \frac{g_1(h) G(h)}{W[g_1,g_2]}dh+C_2
\end{align}
with $C_1$ and $C_2$ are integration constants and $W$ is the Wronskian,
\begin{equation}
W[g_1,g_2]=g_1'g_2-g_2 g_2'.
\end{equation}
It follows that,
\begin{align}
g(h)&=C_1g_1(h)+C_2g_2(h)+\frac{d-\alpha-2}{2(\alpha+1)(2+2\alpha-d)h}
\nonumber\\
&+\frac{h}{(2 ((1 + \alpha) (2 +\alpha) (2 + 2 \alpha - d) (4 + 2 \alpha - d) (d-4) (d-2)^2 (-1 + \cos\Omega)))}
\nonumber\\
&\times \Big(\alpha (d-1) (-2 d^2 + d (8 + (2 +\alpha) n^2) - 2 (4 + (3 + 2 \alpha) n^2) 
\nonumber\\
&+ (8 - 2 d^2 (n-1) - 8 n -  2 (3 + 2 \alpha) n^2 + d (-8 + 8 n + (2 + \alpha) n^2)) \cos\Omega)\Big)+\cdots.
\end{align}
Recall that $g$ was defined in \eqref{eqans} through $\rho$ and the latter should be finite as $h\rightarrow 0$. This can be considered as a boundary condition on $g$ and implies $C_1=0$. We therefore summarize the solution as 
\begin{align}\label{g-solution-odd}
g(h)=\sum_{i=0}\Big[c_{2i}h^{2i+d-2\alpha-3}+b_{2i}h^{2i-1}\Big]
\end{align}
where the first term corresponds to $C_2g_2$. Let us remind that this result is found for generic $\alpha$, while in \eqref{u1-u2} integrations may produce some logarithmic terms for some special values of $\alpha$. To find any log term, let us rewrite integrals of \eqref{u1-u2} as the following formal expansions,
\begin{align} 
u_1&=-\sum_{i=0}\int h^{1+2\alpha}(\alpha_{2i}h^{2i})dh
\nonumber\\
u_2&=\sum_{i=0}\int h^{1+2\alpha-d}(\beta_{2i}h^{2i})dh
\end{align}
Then in the second line, if we choose $2\alpha=d-2j-2$ where $j$ is some nonnegative integer, a logarithmic term shows up as,
\begin{equation}
u_2=\beta_{2j}\log h+\cdots
\end{equation}
This is not the case for $u_1$ since $\alpha$ is supposed to be positive. We therefore modify \eqref{g-solution-odd} as
\begin{align}\label{g-modified-odd}
g(h)=\xi_{\alpha j} b_{2j}\log h+\sum'_{i=0}\Big[c_{2i}h^{2i+d-2\alpha-3}+b_{2i}h^{2i-1}\Big]
\end{align}
where $\xi_{\alpha j}=\delta_{2\alpha,d-2j-2}$ and prime on the summation indicates excluding $b_{2j}$ term.

For even dimensions, we consider $d=4$ for simplicity and regarding $y$ expansion in \eqref{y-even-d}, we expect $g$ includes  logarithmic terms. We therefore suggest the following expansion for the homogeneous equation \eqref{homog},
\begin{align}
g(h)=h^x\Big(c_0+c_1 h +c_2 h^2+c_3 h^3 +\cdots +(\tilde{c}_0+\tilde{c}_2h^2+\cdots)\log h+\cdots\Big)
\end{align}
Plugging into \eqref{homog} gives either $x=x_1=-3-2\alpha$ or $x=x_2=1-2\alpha$ and corresponding solutions respectively are,
\begin{align}
g_1(h)&=h^{-3-2\alpha}\Big[\Big(1+(\alpha^2+\alpha+\frac{5}{16}+\frac{5}{16} \csc ^2\frac{\Omega }{2})h^2+\cdots\Big)+\log h\Big(\frac{(3-\cos\Omega)}{4(1-\cos\Omega)}h^2 \nonumber\\
&+\frac{1}{8} \left(a \left(-8 a^3+7 a+2\right)+a (a+2) \csc ^2(\frac{\Omega }{2})+\csc ^4(\frac{\Omega }{2})-2\right)h^4+\cdots\Big)  \nonumber\\
&+(\log h)^2\Big(\frac{1}{64}  (\cos \Omega-3)^2 \csc ^4(\frac{\Omega }{2})h^4+\cdots\Big)+\cdots\Big],\\
g_2(h)&=h^{1-2\alpha}\Big[\Big(1-\Big(\frac{37}{48}-\alpha+\frac{\alpha^2}{3}-\frac{19}{48}\csc^2\frac{\Omega}{2}\Big)h^2+\cdots\Big)+\log h\Big(\frac{(3-\cos\Omega)}{4(1-\cos\Omega)}h^2 \nonumber\\
&-\frac{1}{768} (\cos \Omega-3) \csc ^4(\frac{\Omega }{2}) ((16 (a-3) a+79) \cos \Omega-16 (a-3) a-29)h^4+\cdots\Big) \nonumber\\
&+(\log h)^2\Big(\frac{1}{64} (\cos \Omega-3)^2 \csc ^4(\frac{\Omega }{2})h^4+\cdots\Big) +\cdots\Big].
\end{align}
Now we apply the same procedure as \eqref{u1g1-u2g2} and \eqref{u1-u2} to find the general solution to the inhomogeneous equation \eqref{inhomog} as follows,
\begin{align}\label{general-g-4d}
g(h)|_{(d=4)}&=C_1g_1(h)+C_2g_2(h)+\frac{-2+\alpha}{4(1-\alpha^2)h}
\nonumber\\
&+\frac{3 h \left(-2 \alpha^3-\alpha^2+2 \alpha-4+\left(2 \alpha^3+11 \alpha^2+10 \alpha-4\right) \csc ^2(\frac{\Omega }{2})\right)}{64 \alpha (\alpha+2)^2 \left(\alpha^2-1\right)}
\nonumber\\
&+\frac{3  (\cos \Omega-3) \csc ^4(\frac{\Omega }{2})}{256 (\alpha^2-1) \alpha (\alpha+2)^2 (\alpha+3)}\log h
\nonumber\\
&\times \Big[-4 \alpha (\alpha+2) (\alpha+3)(1-\cos\Omega)h
\nonumber\\
&+\big(\left(\alpha^4-5 \alpha^3-22 \alpha^2-18 \alpha-3\right)  \cos\Omega+\left(\alpha^4+5 \alpha^3-4 \alpha^2-14 \alpha+9\right)\Big) h^3\Big]
\nonumber\\
&+\frac{3 (\cos\Omega-3)^2 \csc ^4(\frac{\Omega }{2})}{256 (\alpha^2-1) (\alpha+2)} h^3 \log ^2h+\cdots.
\end{align}
Again, we should put $C_1=0$. Similar to odd dimensions, the above result is for a generic $\alpha$ and not valid for $\alpha=1$. In the latter case, we find.
\begin{align}
g(h)|_{(d=4,\alpha=1)}&=-\frac{1}{16 h}+\frac{h \left(95 \csc ^2(\frac{\Omega }{2})-97\right)}{2304}+\cdots
\nonumber\\
&+\log h\Big(
\frac{1}{4 h}+\frac{1}{96} h (\cos\Omega+9) \csc ^2(\frac{\Omega }{2})+\cdots\Big)
\nonumber\\
&+\log^2 h\Big(
\frac{1}{32} h (\cos\Omega-3) \csc ^2(\frac{\Omega }{2})+\cdots\Big)
\nonumber\\
&+\log^3 h\Big(\frac{1}{256} h^3 (\cos\Omega-3)^2 \csc ^4(\frac{\Omega }{2})+\cdots\Big)\cdots.
\end{align}

We summarize the results for $d=4$ as,
\begin{align}\label{general-g-d4}
g(h)|_{(d=4)}&=\sum_{i=0}b_{2i}h^{2i-1}+\log h\sum_{i=0}\tilde{b}_{2i}h^{2i-1}+(\log h)^2\sum_{i=0}\hat{b}_{2i}h^{2i-1}+\cdots.
\end{align}

Then we take $d=6$ and $n=1$. Similar procedure gives,
\begin{align}\label{general-g-6d}
g(h)|_{(d=6)}&=C_1g_1(h)+C_2g_2(h)-\frac{\alpha-4}{4 (\alpha-2) (\alpha+1) h}-\frac{5 \alpha h ((\alpha+3) \cos (\Omega )+\alpha-13)}{128 \left(\alpha^4-5 \alpha^2+4\right) (\cos (\Omega )-1)}
\nonumber\\
&+\frac{45 h^3 \log (h) (116 \cos (\Omega )-15 \cos (2 \Omega )-125) (4 \cos (\Omega )-\cos (2 \Omega )-3) \csc ^8\left(\frac{\Omega }{2}\right)}{65536 (\alpha-2) (\alpha+1) (\alpha+3)}
\nonumber\\
&+\frac{405 h^7 \log ^2(h) (-116 \cos (\Omega )+15 \cos (2 \Omega )+125)^2 \csc ^8\left(\frac{\Omega }{2}\right)}{33554432 (\alpha-2) (\alpha+1) (\alpha+3)}+\cdots.
\end{align}
Here, the exceptional cases are $\alpha=1$ and $\alpha=2$,
\begin{align}\label{general-g-6d-alpha1}
g(h)|_{(d=6,\alpha=1)}&=-\frac{3}{8 h}+h \left(\frac{421 \cos (\Omega )}{2304 (\cos (\Omega )-1)}+\frac{193 \csc ^2\left(\frac{\Omega }{2}\right)}{1536}\right)+\cdots
\nonumber\\
&+\log h\Big(\frac{5}{96} h  (\csc ^2(\frac{\Omega }{2})+1)+\cdots\Big)
\nonumber\\
&-\frac{15 h^5 \log ^2h (961 \cos (\Omega )-206 \cos (2 \Omega )+15 \cos (3 \Omega )-866) \csc ^6\left(\frac{\Omega }{2}\right)}{524288}
\nonumber\\
&+\cdots.
\\
\label{general-g-6d-alpha2}
g(h)|_{(d=6,\alpha=2)}&=-\frac{1}{18 h}+\frac{(269 \cos (\Omega )+61) \csc ^2\left(\frac{\Omega }{2}\right)}{18432}h+\cdots\nonumber\\
&+\log h\Big(\frac{1}{3 h}+\frac{5}{768} h  (5 \cos (\Omega )-11) \csc ^2(\frac{\Omega }{2})+\cdots\Big)
\nonumber\\
&+ \log ^2h\Big(\frac{3 h^3 (-116 \cos (\Omega )+15 \cos (2 \Omega )+125) \csc ^4\left(\frac{\Omega }{2}\right)}{4096}+\cdots\Big)
\end{align}
For $n=2$ in $d=6$ dimension, we find similar results.

So far, we derived solutions to \eqref{g-equation} for $c_n\times R^m$ geometry. The case of $c_n$ can be reached by replacing $m=0$ and $n=d-3$ in the above results. For $k\times R^m$, by similar procedure, we find for odd dimensions,
\begin{align}
g(h)&=C_1\Big(h^{-2\alpha-d-1}-\frac{\left(-4 \alpha^2-2 \alpha-d^2+3 d-2\right)}{2 (d-2)}h^{-2\alpha-d+1}+\cdots\Big) 
\nonumber\\
&+C_2\Big(h^{-2 \alpha-1}-h^{-2 \alpha+1}\frac{\left(2 \alpha^2-2 \alpha d+\alpha-d+1\right)}{d+2}+\cdots\Big) 
\nonumber\\
&-\frac{1}{2 h (2 \alpha+d)}+\frac{(d-1)^2 h}{2 (\alpha+1) (2 \alpha+d) (2 \alpha+d+2)}+\cdots
\end{align}
Of course we should set $C_1=0$. For even dimensions, we consider $d=4$,
\begin{align}
g(h)|_{(d=4)}&=C_2\Big[h^{-2 \alpha-1}+\left(-\frac{\alpha^2}{3}+\frac{7 \alpha}{6}+\frac{1}{2}\right) h^{-2 \alpha+1} +\cdots \nonumber\\
&+\log h \Big(\frac{1}{24} \left(8 \alpha^6-36 \alpha^5+14 \alpha^4+21 \alpha^3-4 \alpha^2-3 \alpha\right) h^{-2 \alpha+5}+\cdots\Big)\Big]
\nonumber\\
&+\frac{1}{4 (\alpha+2) h}-\frac{9 h}{8 (\alpha+1) (\alpha+2) (\alpha+3)}+\cdots\nonumber\\
&-\log h\Big(\frac{ \alpha^2 \left(4 \alpha^3-4 \alpha^2-\alpha+1\right) }{6144 (\alpha+2) (\alpha+5) (\alpha+7)}\nonumber\\
&\times \left(8 \alpha^6-132 a^5+710 \alpha^4-1515 \alpha^3+1082 \alpha^2+297 \alpha-288 K_d^2-270\right)h^9+\cdots\Big)
\end{align}
Therefore the formal expansion of g can be written as,
\begin{align}\label{kRm-g}
g(h)&=\sum_{i=0}e_{2i}h^{2i-2\alpha-1}+\sum_{i=0}h^{2i-1}\Big[d_{2i}+\epsilon_d(\tilde{d}_{2i}\log h+\cdots)\Big].
\end{align}
in which several coefficients may be vanishing. 

\section{Integrands Expansions}\label{B}
In this appendix, we find asymptotic series in \eqref{series1} for integrands of the entanglement entropy. It can be done simply by substituting,
\begin{align}
&\dot{h}=\frac{\sqrt{1-y^{2}}}{\dot{y}(h)}   \nonumber\\
&\sin(\theta)=y(h)
\end{align}
and using the series solution for $y$ in \eqref{Y} and \eqref{y-even-d} in odd and even dimensions, respectively.
For odd dimensions, we have,
\begin{align}\label{app-series1}
\frac{\sin^{n}(\theta)\sqrt{1+h^{2}+\dot{h}^{2}}}{\dot{h}h^{n+2}}&\sim \sum_{i=0}P_{n-2i+2}\frac{1}{h^{n-2i+2}}
\end{align}
with 
\begin{align} 
P_{n+2}=&-\sin^n(\frac{\Omega}{2}),\qquad P_{n}=\frac{n^2(d-3)}{8(d-2)^2}\sin^{n-4}(\frac{\Omega}{2})\sin^2(\Omega),
\nonumber\\
P_{n-2}=&-\frac{n^2}{16(d-4)(d-2)^4}\Big[d^3 (2 - 2 n + n^2) - 2 d^2 (6 - 8 n + 5 n^2)
- 8 (2 - 4 n + 5 n^2) 
\nonumber\\
& + d (24 - 40 n + 33 n^2) + (d^2 (12 + 20 n - 10 n^2) +8 (2 + 6 n - 5 n^2)
\nonumber\\
& + d^3 (-2 - 2 n + n^2) + 
d (-24 - 56 n + 33 n^2)) \cos(\Omega)\Big] \cos^2(\frac{\Omega}{2})\sin^{n-4}(\frac{\Omega}{2})
\end{align}
Similarly,
\begin{align} 
\frac{-\sin^{n}(\theta)(h^{2}+\dot{h}^{2})}{\dot{h}h^{n+2}\sqrt{1+h^{2}+\dot{h}^{2}}}&\sim
\sum_{i=0}Q_{n-2i+2}\frac{1}{h^{n-2i+2}}
\end{align}
where 
\begin{align} 
Q_{n+2}=&\sin^n(\frac{\Omega}{2}),\qquad
Q_{n}=\frac{(d-1)n^2}{8(d-2)^2}\sin^{n-4}(\frac{\Omega}{2})\sin^2(\Omega), \nonumber\\
Q_{n-2}=&\frac{n^2}{16 (d-4) (d-2)^4}  \Big(-8 (10 - 4 n + n^2) + d^3 (2 - 2 n + n^2) - 2 d^2 (14 - 8 n + 3 n^2)
\nonumber\\
& + d (88 - 40 n + 13 n^2) + (d^2 (28 + 4 n - 6 n^2) + d^3 (-2 - 2 n + n^2)
\nonumber\\
& - 8 (-10 + 2 n + n^2) + d (-88 + 8 n + 13 n^2)) \cos\Omega\Big)\cos^2(\frac{\Omega}{2}) \sin^{n-4}(\frac{\Omega}{2})
\end{align}
and 
\begin{align} 
\frac{-\sin^{n}(\theta)\dot{h}g}{h^{n}\sqrt{1+h^{2}+\dot{h}^{2}}}&\sim g(h)
\sum_{i=0}M_{n-2i}\frac{1}{h^{n-2i}}
\end{align}
where 
\begin{align} 
M_{n}=&\sin^n(\frac{\Omega}{2}), \nonumber\\
M_{n-2}=&\frac{(d-1) n^2 \sin ^2(\Omega ) \sin ^{n-4}\left(\frac{\Omega }{2}\right)}{8 (d-2)^2}, \nonumber\\
M_{n-4}=&\frac{n^2}{16 (d-4) (d-2)^4}  \Big(d^2 (36 + 16 n - 6 n^2) - 8 (-6 - 4 n + n^2) + d^3 (-6 - 2 n + n^2) 
\nonumber\\
&+ d (-72 - 40 n + 13 n^2) + (d^2 (-36 + 4 n - 6 n^2) + 
    d^3 (6 - 2 n + n^2) 
\nonumber\\    
& - 8 (6 + 2 n + n^2) + 
    d (72 + 8 n + 13 n^2)) \cos\Omega\Big)\cos^2(\frac{\Omega}{2}) \sin^{n-4}(\frac{\Omega}{2})
\end{align}

\begin{align} 
\frac{\sin^{n}(\theta)g}{\dot{h}h^{n+1}\sqrt{1+h^{2}+\dot{h}^{2}}}&\sim 
g(h)\sum_{i=0}N_{n-2i-1}\frac{1}{h^{n-2i-1}}
\end{align}
\begin{align} 
N_{n-1}=&\frac{n^2}{4(d-2)^2}\sin^{n-4}(\frac{\Omega}{2})\sin^2(\Omega), \nonumber\\
N_{n-3}=&\frac{-n^2}{4 (d-4) (d-2)^4}  \Big(d (16 - 5 n^2) + d^2 (-4 + n^2) + 
  8 (-2 + n^2) + (d^2 (-2 + n)^2 
\nonumber\\
&+ d (-16 + 16 n - 5 n^2) + 8 (2 - 2 n + n^2))\cos\Omega \Big)\cos^2(\frac{\Omega}{2}) \sin^{n-4}(\frac{\Omega}{2})
\end{align}

 In even dimensions, we need to include logarithmic terms in the expansions. Let us demonstrate it for $d=4$,
\begin{align}\label{P}
\frac{\sin^{n}(\theta)\sqrt{1+h^{2}+\dot{h}^{2}}}{\dot{h}h^{n+2}}&\sim \sum_{i=0}\Big(P_{n-2i+2}+\tilde{P}_{n-2i+2}\log h+\hat{P}_{n-2i+2}\log^2h+\cdots\Big)\frac{1}{h^{n-2i+2}}
\end{align}
with taking $d=4$ and $n=1$,
\begin{align} 
P_{3}=&-\sin(\frac{\Omega}{2}),\qquad P_{1}=\frac{1}{8} \cos (\frac{\Omega }{2}) \cot (\frac{\Omega }{2}),
\nonumber\\
P_{-1}=&\frac{1}{256} (11 \cos\Omega +3) \cot ^2(\frac{\Omega }{2}) \csc (\frac{\Omega }{2}),
\nonumber\\
\tilde{P}_{-1}=& -\frac{1}{64} (\cos\Omega -3) \cot ^2(\frac{\Omega }{2}) \csc (\frac{\Omega }{2}),
\nonumber\\
P_{-3}=&\frac{1}{147456}(836 \cos\Omega +469 \cos (2 \Omega )+79) \cot ^2(\frac{\Omega }{2}) \csc ^3(\frac{\Omega }{2}),
\nonumber\\
\tilde{P}_{-3}=&\frac{-1}{1536}(-16 \cos\Omega +\cos (2 \Omega )+31) \cot ^2(\frac{\Omega }{2}) \csc ^3(\frac{\Omega }{2}),
\nonumber\\
\hat{P}_{-3}=&\frac{-1}{512}(\cos\Omega -3)^2 \cot ^2(\frac{\Omega }{2}) \csc ^3(\frac{\Omega }{2}).
\end{align}
Similarly,
\begin{align}\label{Q}
\frac{-\sin^{n}(\theta)(h^{2}+\dot{h}^{2})}{\dot{h}h^{n+2}\sqrt{1+h^{2}+\dot{h}^{2}}}&\sim
\sum_{i=0}\Big(Q_{n-2i+2}+\tilde{Q}_{n-2i+2}\log h+\hat{Q}_{n-2i+2}\log^2h+\cdots\Big)\frac{1}{h^{n-2i+2}}
\end{align}
where 
\begin{align} 
Q_{3}=&\sin(\frac{\Omega}{2}),\qquad Q_{1}=-\frac{3}{8} \cos (\frac{\Omega }{2}) \cot (\frac{\Omega }{2}),
\nonumber\\
Q_{-1}=&\frac{1}{256} (37-19 \cos\Omega ) \cot ^2(\frac{\Omega }{2}) \csc (\frac{\Omega }{2}),
\nonumber\\
\tilde{Q}_{-1}=& \frac{1}{64} (-3) (\cos\Omega -3) \cot ^2(\frac{\Omega }{2}) \csc (\frac{\Omega }{2}),
\end{align}
 
\begin{align} 
Q_{-3}=&-\frac{1}{147456}(-17092 \cos\Omega +811 \cos (2 \Omega )+14641) \cot ^2(\frac{\Omega }{2}) \csc ^3(\frac{\Omega }{2}),
\nonumber\\
\tilde{Q}_{-3}=&-\frac{1}{384} (-65 \cos\Omega +8 \cos (2 \Omega )+59) \cot ^2(\frac{\Omega }{2}) \csc ^3(\frac{\Omega }{2}),
\nonumber\\
\hat{Q}_{-3}=&-\frac{1}{512} (\cos\Omega -3)^2 \cot ^2(\frac{\Omega }{2}) \csc ^3(\frac{\Omega }{2}).
\end{align}
and
\begin{align} \label{M}
\frac{-\sin^{n}(\theta)\dot{h}g}{h^{n}\sqrt{1+h^{2}+\dot{h}^{2}}}&\sim 
g(h)\sum_{i=0}\Big(M_{n-2i}+\tilde{M}_{n-2i}\log h+\hat{M}_{n-2i}\log^2h+\cdots\Big)\frac{1}{h^{n-2i}}
\end{align}
where 
\begin{align} 
M_{1}=&\sin(\frac{\Omega}{2}),\qquad
M_{-1}=\frac{1}{8} (-3) \cos (\frac{\Omega }{2}) \cot (\frac{\Omega }{2}),
 \nonumber\\
M_{-3}=&\frac{1}{256} (13 \cos\Omega +5) \cot ^2(\frac{\Omega }{2}) \csc (\frac{\Omega }{2}),
\nonumber\\
\tilde{M}_{-3}=& -\frac{3}{64} (\cos\Omega -3) \cot ^2(\frac{\Omega }{2}) \csc (\frac{\Omega }{2}),
\nonumber\\
M_{-5}=&\frac{1}{147456}(3268 \cos\Omega +341 \cos (2 \Omega )-1969) \cot ^2(\frac{\Omega }{2}) \csc ^3(\frac{\Omega }{2}),
\nonumber\\
\tilde{M}_{-5}=&-\frac{1}{384} (-17 \cos\Omega +2 \cos (2 \Omega )+17) \cot ^2(\frac{\Omega }{2}) \csc ^3(\frac{\Omega }{2}),
\nonumber\\
\hat{M}_{-5}=&-\frac{ \left(\cos \left(\frac{3 \Omega }{2}\right)-5 \cos (\frac{\Omega }{2})\right)^2 \csc ^5(\frac{\Omega }{2})}{2048}.
\end{align}

\begin{align} \label{N}
\frac{\sin^{n}(\theta)g}{\dot{h}h^{n+1}\sqrt{1+h^{2}+\dot{h}^{2}}}&\sim 
g(h)\sum_{i=0}\Big(N_{n-2i-1}+\tilde{N}_{n-2i-1}\log h+\hat{N}_{n-2i-1}\log^2h+\cdots\Big)\frac{1}{h^{n-2i-1}},
\end{align}
\begin{align} 
N_{0}=&\frac{1}{4} \cos (\frac{\Omega }{2}) \cot (\frac{\Omega }{2}), 
\nonumber\\
N_{-2}=&\frac{1}{32} (\cos\Omega -5) \cot ^2(\frac{\Omega }{2}) \csc (\frac{\Omega }{2}),
\nonumber\\
\tilde{N}_{-2}=&\frac{1}{16} (\cos\Omega -3) \cot ^2(\frac{\Omega }{2}) \csc (\frac{\Omega }{2}),
\nonumber\\
N_{-4}=&\frac{1}{8192}(-996 \cos\Omega +19 \cos (2 \Omega )+809) \cot ^2(\frac{\Omega }{2}) \csc ^3(\frac{\Omega }{2}),
\nonumber\\
\tilde{N}_{-4}=&\frac{1}{512} (-92 \cos\Omega +11 \cos (2 \Omega )+89) \cot ^2(\frac{\Omega }{2}) \csc ^3(\frac{\Omega }{2}),
\nonumber\\
\hat{N}_{-4}=&\frac{1}{1024} \left(\cos (\frac{3 \Omega }{2})-5 \cos (\frac{\Omega }{2})\right)^2 \csc ^5(\frac{\Omega }{2}).
\end{align}

It is worth mentioning that in each expansion, the coefficients of terms with $h$ degrees lower than the first log term are the same in odd and even dimensions. Also note that  nonvanishing coefficients of $log$ and $log^2$,  in the series expansions in \eqref{P}, \eqref{Q}, \eqref{M} and \eqref{N}, start from $i=2$ and $i=3$, respectively.

Similar calculation in $d=6$ with $n=1$ shows that 
\begin{align}
\tilde{P}_i&=0 \quad \text{for}\quad  i>-3 ,\qquad \hat{P}_i=0 \quad \text{for}\quad  i>-7  \\
\tilde{Q}_i&=0 \quad \text{for}\quad  i>-3,\qquad \hat{Q}_i=0 \quad \text{for}\quad  i>-7  \\
\tilde{M}_i&=0 \quad \text{for}\quad  i>-5,\qquad \hat{M}_i=0 \quad \text{for}\quad  i>-9  \\
\tilde{N}_i&=0 \quad \text{for}\quad  i>-4,\qquad \hat{N}_i=0 \quad \text{for}\quad  i>-8 
\end{align}



\end{document}